%% file: main.tex
\documentclass[sigconf]{acmart}

\AtBeginDocument{%
  \providecommand\BibTeX{{%
    \normalfont B\kern-0.5em{\scshape i\kern-0.25em b}\kern-0.8em\TeX}}}

\usepackage{tabularx}
\usepackage{enumitem}
\usepackage{xspace}
\usepackage{appendix}

\copyrightyear{2024}
\acmYear{2024}
\setcopyright{rightsretained}
\acmConference[ASSETS '24]{The 26th International ACM SIGACCESS Conference on Computers and Accessibility}{October 27--30, 2024}{St. John's, NL, Canada}
\acmBooktitle{The 26th International ACM SIGACCESS Conference on Computers and Accessibility (ASSETS '24), October 27--30, 2024, St. John's, NL, Canada}
\acmDOI{10.1145/3663548.3675634}
\acmISBN{979-8-4007-0677-6/24/10}

\begin{document}

\input{commands}
%%
%% The "title" command has an optional parameter,
%% allowing the author to define a "short title" to be used in page headers.
\title{Uncovering the New Accessibility Crisis in Scholarly PDFs}
% \title{A New Crisis in Scholarly PDF Accessibility}
\subtitle{Publishing Model and Platform Changes Contribute to Declining Scholarly Document Accessibility in the Last Decade}
%%
%% The "author" command and its associated commands are used to define
%% the authors and their affiliations.
%% Of note is the shared affiliation of the first two authors, and the
%% "authornote" and "authornotemark" commands
%% used to denote shared contribution to the research.

\author{Anukriti Kumar}
\email{anukumar@uw.edu}
\orcid{}
\affiliation{%
  \institution{University of Washington}
%   \streetaddress{1410 NE Campus Parkway}
  \city{Seattle}
  \state{WA}
  \postcode{98103}
  \country{USA}
}

\author{Lucy Lu Wang}
\email{lucylw@uw.edu}
\orcid{0000-0001-8752-6635}
\affiliation{%
  \institution{University of Washington}
%   \streetaddress{1410 NE Campus Parkway}
  \city{Seattle}
  \state{WA}
  \postcode{98103}
  \country{USA}
}

%%
%% By default, the full list of authors will be used in the page
%% headers. Often, this list is too long, and will overlap
%% other information printed in the page headers. This command allows
%% the author to define a more concise list
%% of authors' names for this purpose.
\renewcommand{\shortauthors}{Anukriti Kumar and Lucy Lu Wang}
%%
%% The abstract is a short summary of the work to be presented in the
%% article.
\begin{abstract}
Most scholarly works are distributed online in PDF format, which can present significant accessibility challenges for blind and low-vision readers. To characterize the scope of this issue, we perform a large-scale analysis of 20K open- and closed-access scholarly PDFs published between 2014--2023 sampled across broad fields of study. We assess the accessibility compliance of these documents based on six criteria: Default Language, Appropriate Nesting, Tagged PDF, Table Headers, Tab Order, and Alt-Text; selected based on prior work and the SIGACCESS Guide for Accessible PDFs \cite{Trewin_2014}. To ensure robustness, we corroborate our findings through automated accessibility checking, manual evaluation of alt text, comparative assessments with an alternate accessibility checker, and manual assessments with screen readers. Our findings reveal that less than 3.2\% of tested PDFs satisfy all criteria, while a large majority (74.9\%) fail to meet any criteria at all. Worse yet, we observe a concerning drop in PDF accessibility since 2019, largely among open access papers, suggesting that efforts to improve document accessibility have not taken hold and are on a backslide. While investigating factors contributing to this drop, we identify key associations between fields of study, creation platforms used, models of publishing, and PDF accessibility compliance, suggesting that publisher and author choices significantly influence document accessibility. This paper highlights a new crisis in scholarly document accessibility and the need for a multi-faceted approach to address the problem, involving the development of better tools, enhanced author education, and systemic changes in academic publishing practices.

  % The majority of scientific papers are distributed in PDF, which pose accessibility challenges for blind and low vision readers. <> We characterize the scope of this problem by performing a large-scale analysis of 20K open and closed access scholarly PDFs published in the years 2014--2023 sampled across various fields of study. We began by studying the accessibility compliance of these documents over six criteria: Default language, Appropriate nesting, Tagged PDF, Table headers, Tab order, and Alt-Text. Less than 3.2\% of these PDFs satisfy all the criteria, while a large majority of PDFs (74.86\%) in our sample met 0 criteria. There was also variability in compliance across different fields of study, and we investigated factors that may have contributed to this discrepancy. We find associations between certain typesetting software, models of publishing and the resulting PDF accessibility compliance, indicating that publisher or author choices around typesetting tools and access choice can have unintended consequences for accessibility. We further explore manual assessments with screen readers (NVDA and VoiceOver) and observe that in many cases, automated accessibility checkers fail to capture the nuances necessary for true accessibility, underscoring the need for manual evaluations and enhanced checker capabilities. This paper calls for a multi-faceted approach to improve accessibility, involving better tools, author education, and systemic changes in academic publishing practices.
\end{abstract}
\maketitle
%%
%% The code below is generated by the tool at http://dl.acm.org/ccs.cfm.
%% Please copy and paste the code instead of the example below.
%%
\begin{CCSXML}
<ccs2012>
   <concept>
       <concept_id>10003120.10011738.10011773</concept_id>
       <concept_desc>Human-centered computing~Empirical studies in accessibility</concept_desc>
       <concept_significance>500</concept_significance>
       </concept>
   <concept>
       <concept_id>10003120.10011738.10011774</concept_id>
       <concept_desc>Human-centered computing~Accessibility design and evaluation methods</concept_desc>
       <concept_significance>500</concept_significance>
       </concept>
 </ccs2012>
\end{CCSXML}

\ccsdesc[500]{Human-centered computing~Empirical studies in accessibility}
\ccsdesc[500]{Human-centered computing~Accessibility design and evaluation methods}

%%
%% Keywords. The author(s) should pick words that accurately describe
%% the work being presented. Separate the keywords with commas.
\keywords{accessibility, document accessibility, scholarly documents, blind and low vision, science of science}

\hyphenation{Open-Alex Py-PDF}

%% A "teaser" image appears between the author and affiliation
%% information and the body of the document, and typically spans the
%% page.

% \received{20 February 2007}
% \received[revised]{12 March 2009}
% \received[accepted]{5 June 2009}

%%
%% This command processes the author and affiliation and title
%% information and builds the first part of the formatted document.
\input{0_introduction}

\input{1_related_work}
\input{2_sos_methods}

\input{3_sos_results}

\input{4_discussion}

\section*{Acknowledgments} This work is partially supported by the University of Washington eScience Insitute's Azure Cloud Credits for Research and Teaching.

\bibliographystyle{ACM-Reference-Format}
\bibliography{A11y, lucy}

\appendix
\input{appendix}

\end{document}

%% file: commands.tex
% commenting
\newcommand\lucy[1]{{\color{blue}\{\textit{#1}\}$_{lucy}$}}

\newcommand\todoit[1]{{\color{red}\{TODO: \textit{#1}\}}}
\newcommand\todo{{\color{red}{TODO}}\xspace}
\newcommand\todocite{{\color{red}{CITE}}\xspace}

\newcommand\uwname{{our institution\xspace}}
% \newcommand\uwname{the University of Washington\xspace}

% special table column type (p but raggedright); use in conjunction with X
\newcolumntype{L}[1]{>{\raggedright\let\newline\\\arraybackslash\hspace{0pt}}p{#1}}
\newcolumntype{M}[1]{>{\raggedright\let\newline\\\arraybackslash\hspace{0pt}}m{#1}}
\newcolumntype{C}[1]{>{\centering\arraybackslash\hspace{0pt}}p{#1}}

% vertical rule
\newcommand{\rulesep}{\unskip\ \textcolor{gray}{\vrule}\ }

% colors
\definecolor{darkgreen}{rgb}{0.0, 0.4, 0.13}

%% file: 0_introduction.tex
\section{Introduction}
\label{sec:introduction}

% Our goals are (i) to describe the current state of scholarly PDF accessibility and how circumstances have changed over time, (ii) understand the limitations of accessibility checker software, and (iii) provide recommendations for producing and checking the accessibility of scholarly PDFs.  

Although the proliferation of digital documents in academic publishing has significantly increased the availability of content, the true accessibility of these documents still remains a significant challenge, especially for blind and low-vision (BLV) readers or screen-reader users. Despite advancements in digital publishing technologies, including document creation software, automated accessibility checkers, and evolving standards like the Web Content Accessibility Guidelines (WCAG) and PDF/Universal Accessibility (PDF/UA) guidelines, actual implementations often lag behind these advancements due to significant gaps in the widespread adoption and application of these technologies across the academic landscape.
% Scholarly literature is most commonly available in the form of PDFs, and t
To make PDFs (the most common format for scholarly literature) accessible, they must be annotated with proper reading order, headings, tags, table structure, image alt-text, and more. Unfortunately, creating these annotations is laborious, requiring proprietary tools and a significant amount of knowledge and motivation from authors. Moreover, each PDF variant, even minor revisions, must be annotated anew, posing a continuous barrier to accessibility. As a result, the vast majority of scholarly PDFs are not accessible, leading to high cognitive load and frustration for BLV researchers trying to read these papers. 

\begin{figure}[t!]
  \centering
    \includegraphics[width=0.95\linewidth,trim={0mm 6mm 0mm -6mm},clip]{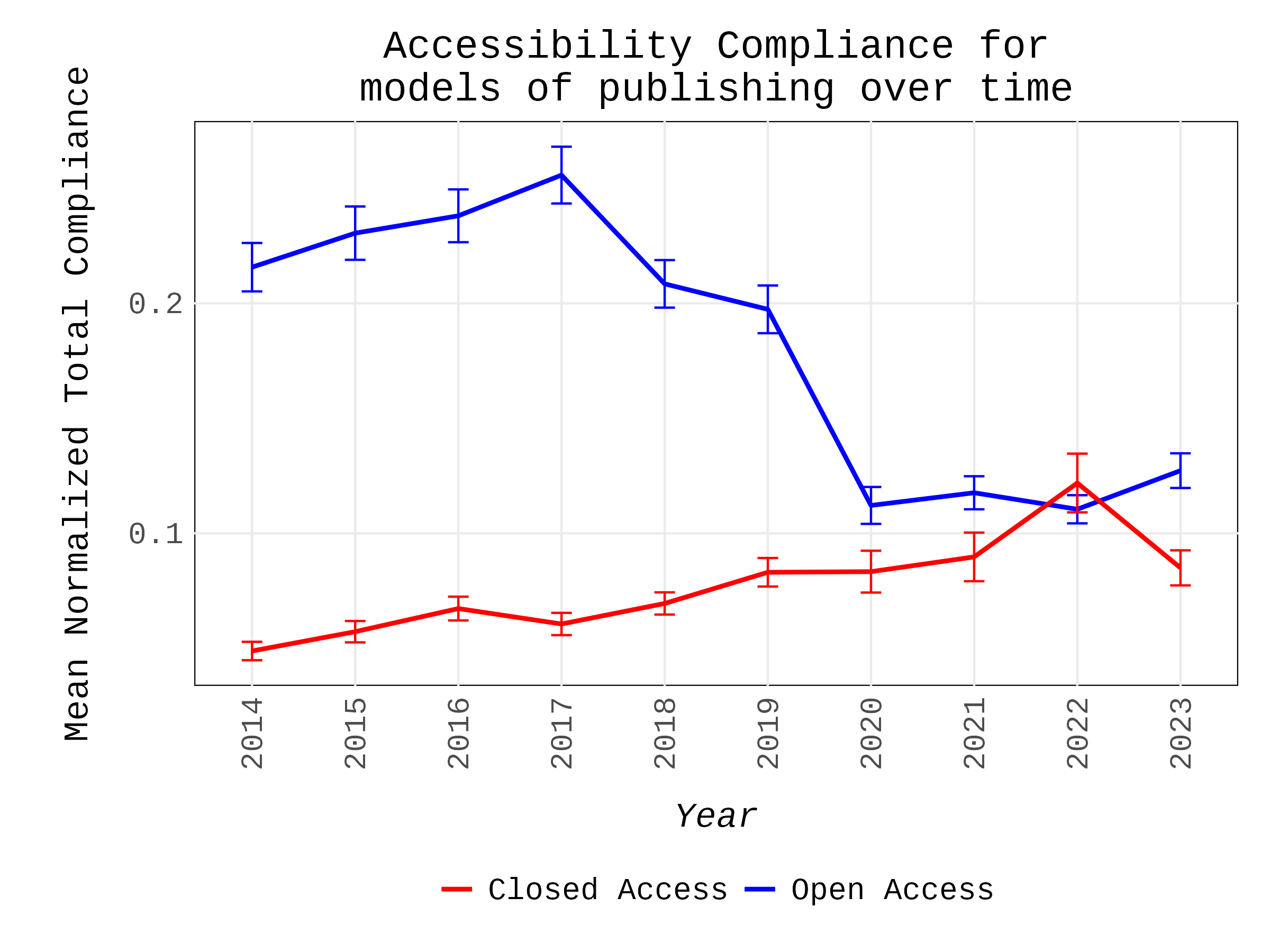}
  \vspace{-1mm}
  \caption{
  Accessibility compliance for models of publishing over time (2014-2023) with error bars ($\pm$ standard error). There is a large decline in mean normalized total compliance after 2019 for open access papers. Following this, the compliance values remain stable or increase slightly.}
  \Description{A line plot displays the mean normalized total compliance of scholarly PDFs across publishing models (open versus closed-access) from 2014 to 2023, with error bars showing standard deviation. The x-axis denotes the year, while the y-axis shows the proportion of PDFs. Open access papers show a significant decline in compliance after 2019, stabilizing or slightly increasing thereafter. Conversely, compliance for closed-access papers has been gradually improving since 2014.}
  \label{fig:compliance_model_time_errors}
\end{figure}

% Recent advancements in digital publishing technologies, including document creation software, automated accessibility checkers, and evolving standards like the Web Content Accessibility Guidelines (WCAG) and PDF/Universal Accessibility (PDF/UA) aim to enhance the accessibility of academic content. However, despite these efforts, actual implementations often lag behind these advancements, resulting from significant gaps in the widespread adoption and application of these technologies across the academic landscape. 

Poor paper PDF accessibility has been documented in prior work \citep{Bigham2016AnUT, Brady2015CreatingAP, Lazar2017MakingTF, Ribera2019PublishingAP, Nganji2015ThePD}, though these studies are limited in terms of generalizability and mechanistic understanding of how technical and publishing decisions contribute to low document accessibility. For example, most prior work analyzed papers published in fields adjacent to accessible computing---human-computer interaction, disability studies, etc---and do not necessarily generalize to other fields of study. The use of automated checkers to evaluate PDF accessibility focuses on technical compliance with accessibility standards rather than the user experience, which also constrains the generalization of results \cite{Brady2015CreatingAP, Lazar2017MakingTF, Ribera2019PublishingAP, Nganji2015ThePD, swiss_2023}. Furthermore, existing research does not explore important questions regarding 
% PDF accessibility, such as 
the impact of various document creation software, publishing models (open access vs. closed access), and field of study on document accessibility.
% across different fields. 
Our work aims to fill these gaps by addressing critical research questions: (i) What are the overall status and trends in scholarly document accessibility? (ii) What are the associations between PDF creation platforms, models of publishing (open versus closed access), fields of study, and scholarly PDF accessibility? (iii) To what extent do automated accessibility checkers reflect the performance of screen readers when reading scholarly documents? 

% \lucy{I edited this last RQ to pull back from BLV user experiences, since we don't conduct such a study}
% To what extent do current automated accessibility checkers reflect the actual accessibility experiences of BLV users when interacting with scholarly documents?

% Furthermore, while automated accessibility checking tools offer a convenient way to assess PDF accessibility, they have inherent limitations that constrain the generalization of their results. These tools primarily focus on technical compliance with accessibility standards rather than the user experience, and hence, do not always capture the full spectrum of accessibility barriers experienced by BLV readers.

% Furthermore, the results obtained from automated accessibility checking tools can not be fully generalized. While these tools offer a convenient way to assess PDF accessibility, they have several inherent limitations. Primarily, they focus on technical compliance with accessibility standards rather than the user experience. As a result, they may not always capture the full spectrum of accessibility barriers experienced by blind or low-vision readers.
% \lucy{would also be good to integrate something here regarding the limitations of accessibility checking tools and why we can't fully generalize from their results}
% , have not been comprehensively examined. 
To answer these questions, we create a corpus of 20K papers published in the last decade (2014-2023), representing scholarly publications across 19 fields and 766 sub-fields of study. Using this corpus, we characterize the current state of PDF accessibility within academic literature across disciplines and publishing models, as well as accessibility trends over time, providing a comprehensive analysis that has been lacking in existing literature. 
Our contributions in this work are summarized below:

\begin{itemize}[topsep=2pt,itemsep=2pt, leftmargin=10pt]
    \item We characterize the state of scientific paper PDF accessibility by estimating the degree of adherence to accessibility criteria for our corpus of 20K papers. Our findings highlight that overall compliance levels are very low (only 3.2\% satisfy all criteria we measure) and highly variable across fields of study. 
    % (ranging from around 6.5\% at the high end for fields like Environmental Science and Medicine to less than 2\% at the low end for Geology, Mathematics, and Physics). 
    Additionally, we identify a concerning trend that overall accessibility has been declining since 2019 and remains lower than before. Despite efforts to address accessibility issues, this finding suggests that existing approaches may not be systematically addressing the underlying challenges, and there is a need to understand the factors contributing to this decline.

% Additionally, we find overall accessibility has been declining since around 2017 and remains lower than before, indicating a concerning trend. Despite efforts to address accessibility issues, this finding indicates that overall compliance levels have not improved and instead, remain lower than in earlier years.

% This raises concerns about the effectiveness of current strategies and initiatives aimed at improving accessibility in academic publishing. It suggests that existing approaches may not be adequately addressing the underlying challenges or that new barriers are emerging over time. Understanding the factors contributing to this decline is essential for developing targeted interventions to reverse the trend and enhance accessibility.
% \lucy{Don't characterize it as slightly increasing here. Focus on the dip that was identified starting in around 2018-2019 and how things do not seem to be improving and that overall accessibility is lower than before. That's the surprising finding here. Maybe combine this last finding into bullet point too since bullet two contains the mechanistic reason for this change.}

    \item Building upon the methods introduced in \citet{Wang2021ImprovingTA}, we study the relationships between creation platforms, publishing models, and fields of study with PDF accessibility, to explain the differences in accessibility compliance we observe.
    % observed across different academic fields over time. 
    Our analysis reveals significant correlations between the creation platform and PDF accessibility, as well as publishing modality and PDF accessibility. Specifically, we find that documents created by some publishers/publishing software (e.g., Microsoft Word, Adobe InDesign, Elsevier) tend to exhibit higher accessibility compliance, while those created by other software (e.g., LaTeX, Arbortext APP, XPP) show low levels of compliance.
    We also observe that while open-access papers exhibit significantly higher accessibility compliance than closed-access papers, that a decline in open-access accessibility accounts for much of the overall decline observed since 2019 (Figure~\ref{fig:compliance_model_time_errors}). 

% Building on these findings, we propose policy and practice recommendations for authors, developers, and publishers. These recommendations include advocating for best practices in typesetting, enhancing alt-text descriptiveness, and implementing strategies to improve compliance with accessibility standards, particularly within the framework of open-access publishing.

    % We find strong correlations between typesetting software and PDF accessibility, with Microsoft Word and LaTeX being positively and negatively correlated respectively with the accessibility compliance. 
    % We further observed open-access papers to be more accessible across all fields of study. Based on these findings, we further propose policy and practice recommendations for authors, developers and publishers, including best practices for typesetting, alt-text descriptiveness, and strategies for improving compliance with accessibility standards, including open-access publishing.
    % , and having significant impact over the accessibility compliance across all fields of study, over time. Based on these findings, we further propose policy and practice recommendations for authors, developers and publishers, including best practices for typesetting, alt-text descriptiveness, and strategies for improving compliance with accessibility standards, including open-access publishing.

    \item We assess the effectiveness and limitations of automated accessibility checkers by leveraging secondary assessments with alternate checkers (Adobe versus PAC3) as well as with screen readers (NVDA and VoiceOver). We find that although automated checkers are valuable for large-scale assessments (and make them possible), relying solely on automated reports may give a false sense of progress, as they can fall short of accurately capturing the nuances of document accessibility. For instance, they do not provide an adequate assessment of alt-text compliance, with many papers that pass the criteria often lacking any usable alt-text. 
    % These checkers also fail to indicate if certain elements are missing from the paper, and still return pass or fail for the associated criteria. 
    % Additionally, we note that if tables are not tagged as tables (and interpreted as paragraphs instead), the Table headers criteria may still pass. 
    Our findings underscore the continued need for manual intervention from authors and publishers. We also highlight opportunities to improve automated checker capabilities to better align them with the actual accessibility experiences of end-users, particularly those relying on screen readers.

\end{itemize}

This paper is organized as follows: Following a description of related work in Section \ref{sec:related_work}, Section \ref{sec:methods} outlines our data collection and methodology for conducting the analysis of the current state of academic PDF accessibility. In Section \ref{sec:sos}, we document our analysis results on the accessibility of academic PDFs across various disciplines. Section \ref{sec:sos_evaluate_checkers} presents results from our extended qualitative analysis of automated accessibility checkers, offering insights into their reliability. Section \ref{sec:discussion} discusses the implications of our findings and proposes recommendations for improving PDF accessibility in academic publishing, followed by limitations in Section \ref{sec:limitations}.

%% file: 1_related_work.tex
\section{Background and Related work}
\label{sec:related_work}
Accessibility is essential for individuals with diverse abilities to fully participate in academic and professional endeavors. For those who are blind and low vision, equitable access to scientific literature can level the playing field, allowing these individuals to pursue education, conduct research, and contribute to scholarly discourse with fewer barriers. Moreover, making digital content, including scholarly literature, accessible is a legal obligation in many jurisdictions, governed by laws such as the Americans with Disabilities Act (ADA) \cite{ADA1990} and the European Accessibility Act \cite{European_accessibility_act}.

\subsection{Challenges with PDF Accessibility}
PDF documents, while widely used in academic publishing, present significant challenges due to their inherent limitations in accessibility features \cite{Ribera2008AreAccessible, Lazar2017MakingTF, NielsenPDFStillUnfit}. 
% These include the lack of proper semantic tagging, inadequate alternative text for images, poor color contrast, and more, which are considered in previous work \cite{Angerbauer2022colorvision, Brady2015CreatingAP, Bigham2016AnUT}.
Although they support accessibility features like semantic tagging and alternative text for images, the effective use of these features is often limited by authoring tools and a lack of proactive authoring practices \cite{Angerbauer2022colorvision, Brady2015CreatingAP, Bigham2016AnUT, Wang2021ImprovingTA}. Screen reader users often struggle with accessing PDF documents because many authors have not created files to be accessible \cite{Bigham2016AnUT}, largely due to three factors: (1) the complexity of the PDF file format, which make it less amenable to certain accessibility features, (2) the dearth of tools, especially non-proprietary tools, for creating accessible PDFs, and (3) the dependency on volunteerism from the community with minimal support or enforcement. These limitations emphasize the PDF format’s original intent, to support faithful visual representation of a document for printing, a goal which conflicts with the needs for accessibility.

There are established guidelines and standards, such as the W3C Web Content Accessibility Guidelines (WCAG) \citep{Chisholm2001WebCA, Caldwell2008WebCA}, and ISO 14289-1 or PDF/UA (Universal Accessibility) \cite{PDF_UA1, PDF_UA2}, that provide comprehensive guidance to authors on making individual elements within PDF documents accessible. These standards also detail the technical implementations necessary to enhance PDF accessibility. However, a significant challenge persists due to the complexity of these guidelines and the depth of understanding of accessibility issues required to effectively apply them, hindering their widespread adoption \cite{Kelly2005}. 

For academic publishing, guidelines and policy changes have been introduced over the past decade to ameliorate some of the issues around scholarly PDF accessibility. Professional organizations such as the Association for Computing Machinery (ACM) have encouraged PDF accessibility through standards and writing guidelines\footnote{\href{https://www.acm.org/publications/authors/submissions}{https://www.acm.org/publications/authors/submissions}}. Influential conferences like the ACM CHI Conference on Human Factors in Computing Systems (CHI) and the ACM SIGACCESS Conference on Computers and Accessibility (ASSETS) have released guidelines and mandates for creating accessible submissions.\footnote{See \href{https://sigchi.org/resources/guides-for-authors/accessibility/}{https://sigchi.org/resources/guides-for-authors/accessibility/} and \href{https://www.sigaccess.org/welcome-to-sigaccess/resources/accessible-conference-guide/}{https://sig-access.org/resources/accessible-conference-guide/}} There has also been a growing trend to outsource some of these accessibility tasks to specialized companies.
% , ensuring that academic content complies with accessibility standards.

Despite having access to these resources and guidelines, \citet{AuthorReflections2022} identifies a gap between accessibility guidance and its practical application in academic writing. The authors analyzed content elements such as tables, charts, and images in 330 papers published during 2011-2020 at ASSETS. The analysis focuses on specific design choices that, while maintaining the visual design of the papers, render these elements inaccessible. \citet{Ribera2019PublishingAP} conducted a case study on DSAI 2016 (Software Development and Technologies for Enhancing Accessibility and Fighting Infoexclusion), where submitting authors identified barriers to creating accessible proceedings, including the lack of sufficient tooling and lack of awareness of accessibility. The authors recommended creating a new role in the organizing committee dedicated to accessible publishing. In recent years, some publishers (including Science, Nature, PLoS, and others) now provide HTML reading experiences for their papers, which can dramatically mitigate challenges for BLV researchers; the ACM Digital Library\footnote{\href{https://dl.acm.org/}{https://dl.acm.org/}} provides some publications in HTML format, which is easier to make accessible than PDF~\cite{Graells2007EstudioDL}. ArXiv has also released a beta version of its HTML Papers project, providing HTML for nearly all newly submitted papers on arXiv along with a number of older ones \cite{Frankston2024HTMLPO}. These policy changes have led to improvements in localized communities, but have not been widely adopted by academic publishers and conference organizers. 

Additionally, \citet{Bigham2016AnUT} and \citet{Brady2015CreatingAP} highlighted the scarcity of tools and dependency on volunteerism as major challenges for creating accessible PDFs. Since 2018, however, a variety of tools and frameworks have been developed to address these issues, focusing on automated solutions \cite{Zulfiqar_2020, Wang2021ImprovingTA, Pradhan_2022}, artificial intelligence (AI) integration \cite{Schmitt-Koopmann_2022, Singh_2024, Mack_2021}, and improved authoring practices \cite{Rajkumar2020PDFAO, arXiv_2022}. However, the remediation process is still labor-intensive and unintuitive, especially for those not well-versed in accessibility standards \cite{AuthorReflections2022}. Many content creators are simply unaware of the needs of visually impaired users or the tools available to improve accessibility \cite{Lazar2017MakingTF}. 
In many fields outside of computing, such as Biology and Medicine, versions of record (the final published versions of papers) are produced by publishers from an author's submitted manuscript. This transfer of control over the paper's PDF accessibility from authors to publishers appears to be a step forward. However, publishers rely on specialized providers for formatting and accessibility remediation, a process that incurs significant costs and time, and which may still not lead to accessible PDFs. 
% This combination of tooling and manual labor deter publishers, particularly those with limited resources. 
These options are also not available to everyone: small publishers typically lack the budget for extensive manual or custom processing that larger journals can afford. Similarly, platforms like ArXiv, struggle to implement comprehensive accessibility measures due to the sheer volume and rapid turnover of documents \cite{Frankston2024HTMLPO}.

\subsection{Assessing the state of PDF Accessibility}

\input{figtabs/tab_rw1.tex}

Table~\ref{tab:prior_work} lists previous studies that have analyzed PDF accessibility of academic papers and shows how our study compares. Prior work has focused on papers published in venues for accessibility, human-computer interaction, disability studies, and related fields, and many studies are specific to certain publication venues. 
% Our analysis tries to quantify paper accessibility more broadly across fields of study. 
\citet{Brady2015CreatingAP} quantified the accessibility of 1,811 papers from CHI 2010-2016, ASSETS 2014, and W4A, assessing the presence of document tags, headers, and language. They found that compliance improved over time as a response to conference organizers offering to make papers accessible as a service to any author upon request. 
\citet{Lazar2017MakingTF} conducted a study quantifying accessibility compliance at CHI from 2010 to 2016 as well as ASSETS 2015, confirming the results of \citet{Brady2015CreatingAP}. They found that across 5 accessibility criteria, the rate of compliance was less than 30\% for CHI papers in each of the 7 years that were studied. The study also analyzed papers from ASSETS 2015, an ACM conference explicitly focused on accessibility, and found that those papers had significantly higher rates of compliance, with over 90\% of papers being tagged for correct reading order and no criteria having less than 50\% compliance. This finding indicates that community buy-in is an important contributor to paper accessibility.

\citet{Nganji2015ThePD} examined 200 PDFs of papers published between 2009 and 2013 from four journals in disability studies, revealing a compliance rate of only 15-30\% for accessibility features like tagging and alternative text for images. In a more extensive analysis \citep{Nganji2018ThePD}, only 15.5\% of the PDF documents were found to be tagged, and just 10.5\% provided alternative text for images, highlighting a significant gap in adherence to the PDF/UA (Universal Accessibility) standards and WCAG 2.0 guidelines.
In a broader survey conducted with 2500 open-access PDF documents from online repositories in Switzerland from 2018-2022 \cite{swiss_2023}, it was revealed that fewer than 11\% of documents were found to have minimal accessibility features. This study identified a lack of knowledge and prioritization of PDF accessibility among repository managers, indicating a broader systemic issue that extends beyond individual publishers or conferences to the infrastructure of academic knowledge storage itself. 
To date, the most extensive study \cite{Wang2021ImprovingTA} analyzed over 11K PDFs from various fields, also concluding that adherence to accessibility criteria remains low, nearly 2.4\% of the documents meeting all accessibility criteria set by researchers, showcasing vast room for improvement in scientific document accessibility.

While these studies provide valuable insights, they tend to be limited in scope and do not cover a wide range of academic fields or track long-term trends. A majority of these works focus on specific content elements without a comprehensive assessment of different aspects of PDF accessibility, such as reading order, logical structure, and the effectiveness of alternative text, which can only be evaluated manually or using screen readers. Additionally, only a few studies have investigated trends over time, crucial for evaluating the impact of efforts to improve accessibility.  In light of these gaps, our study aims to conduct a large-scale, multi-field analysis of scientific PDF accessibility over the past decade, offering insights into trends and factors influencing PDF accessibility in scholarly communication.

%% file: figtabs/tab_rw1.tex
\begin{table*}[t!]
\begin{tabular}{p{25mm}p{13mm}p{50mm}p{18mm}p{50mm}}
    \toprule
    \textbf{Prior work} & \textbf{PDFs analyzed} & \textbf{Venues} & \textbf{Year} & \textbf{Accessibility checker} \\
    \midrule
        \citet{Brady2015CreatingAP} & 1,811 & CHI, ASSETS and W4A & 2011--2014 & PDFA Inspector \\ [0.5mm]
        \addlinespace
        \citet{Lazar2017MakingTF} & 465 + 32 & CHI and ASSETS & 2014--2015 & Adobe Acrobat Action Wizard \\ [0.5mm]
        \addlinespace
        \citet{Ribera2019PublishingAP} & 59 & DSAI & 2016 & Adobe PDF Accessibility Checker 2.0 \\ [0.5mm]
        \addlinespace
        \citet{Nganji2015ThePD, Nganji2018ThePD} & 200 & \textit{Disability \& Society}, \textit{Journal of Developmental and Physical Disabilities}, \textit{Journal of Learning Disabilities}, and \textit{Research in Developmental Disabilities} & 2009--2013 & Adobe PDF Accessibility Checker 1.3 \\ [0.6mm]
        \addlinespace
        \citet{swiss_2023} & 2500 & Swiss repositories & 2018-2022 & Adobe PDF Accessibility Checker \\ [0.6mm]
        \citet{Wang2021ImprovingTA} & 11,397 & Venues across 19 fields of study & 2010-2019 & Adobe Acrobat Accessibility Plug-in Version 21.001.20145 \\ [0.6mm]
        \addlinespace
        \textbf{\textit{Our analysis}} & 19,997 & Venues across 19 fields and 766 sub-fields of study & 2014--2023 & Adobe Acrobat Accessibility Plug-in version 24.002.20687 \\
    \bottomrule \\ [-1mm]
\end{tabular}
\caption{Prior work has investigated PDF accessibility for papers published in specific venues such as CHI, ASSETS, W4A, DSAI, or various disability journals. Several of these works were conducted manually, and were limited to a small number of papers, while more thorough analyses were conducted for CHI and ASSETS, two conference venues focused on accessibility and HCI, and open access papers in a Swiss repository. 
The analysis conducted by \citet{Wang2021ImprovingTA} is over broad fields of study and is closest to ours.
Our study expands on this prior work to investigate accessibility trends over 19997 PDFs sampled from across different fields of study in the last decade.}
\label{tab:prior_work}
\end{table*}

%% file: 2_sos_methods.tex
\section{Data \& Methods}
\label{sec:methods}

To capture and better characterize the scope and severity of problems around academic PDF accessibility, we conduct an analysis over a representative sample of scholarly documents using industry-standard accessibility checker tools. This section provides a thorough overview of the methodology employed to perform this analysis. We hope these methods will guide more accurate monitoring of PDF accessibility in the future.

% (e.g., what proportion of papers have accessible PDFs?), whether the state of PDF accessibility is improving over time (e.g., are papers published in 2019 more likely to be accessible than those published in 2010?), and whether the typesetting software used to create a paper is associated with the accessibility of its PDF (e.g., are papers created using Microsoft Word more or less accessible than papers created with other software?).

% \lucy{move to related work} 
% Prior studies on PDF accessibility have been limited to papers from specific publication venues such as CHI, ASSETS, W4A, DSAI, and journals in disabilities research. Notably, these venues are closer to the field of accessible computing, and are consequently more invested in accessibility.\footnote{See submission and accessibility guidelines for ASSETS (\href{https://assets19.sigaccess.org/creating_accessible_pdfs.html}{https://assets19.sigaccess.org/creating\_accessible\_pdfs.html}), CHI (\href{https://chi2021.acm.org/for-authors/presenting/papers/guide-to-an-accessible-submission}{https://chi2021.acm.org/ for-authors/presenting/papers/guide-to-an-accessible-submission}), W4A (\href{http://www.w4a.info/2021/submissions/technical-papers/}{http://www.w4a.info/2021/submissions/technical-papers/}) and DSAI (\href{http://dsai.ws/2020/submissions/}{http://dsai.ws/2020/submissions/}).}  We expand upon this work by investigating accessibility trends across various fields of study and publication venues. Our goal is to characterize the overall state of paper PDF accessibility and identify ongoing challenges to accessibility going forward.
% Further, we introduce analysis methodology that newly considers typesetting software and alt-text information content; we hope these methods will guide more accurate monitoring of PDF accessibility in the future.

\subsection{Corpus Construction}
\label{subsec:corpus}
We constructed and analyzed a corpus of 20K papers published in the last decade (2014-2023), stratified across domains of study.
We began by sampling documents from OpenAlex,\footnote{\href{https://openalex.org/}{https://openalex.org/}} a large, non-profit-operated resource that indexes the metadata of over 250M scholarly works from 250k sources \cite{priem2022openalexfullyopenindexscholarly}. 
% We used the field of study labels and open/closed access labels defined in OpenAlex associated with each paper. 
Initially, we sampled 60K papers including both open and closed-access articles, as indexed by OpenAlex using the `is\_oa' boolean attribute which classifies a paper as open access if there is a freely accessible URL for the full text, without login requirements. 
% to ensure comprehensive coverage of the academic publishing landscape. 
% Notably, two-thirds of these documents were open-access articles, reflecting the growing trend towards more open access in scholarly research. 
For domains, we used the 19 top-level fields of study (e.g., Biology, Computer Science, Sociology) and their associated 766 subfields as defined by Microsoft Academic Graph \citep{msr:mag1, Shen2018AWS}, which is adopted by OpenAlex. For each subfield of study, we randomly sampled papers from the top 20 venues as defined by aggregate citation count over the last decade.

% , as well as including documents without specified venue information, which can represent books and book chapters. 

% We matched the metadata from OpenAlex to paper PDFs using the Semantic Scholar API \citep{Ammar2018ConstructionOT, Kinney2023TheSS},
% However, after sourcing PDFs through web scraping and the Semantic Scholar literature corpus \citep{Ammar2018ConstructionOT}, we encountered a significant reduction in our dataset.
% identifying PDFs for 35K of the 60K papers.
We downloaded open access PDFs using URLs provided by OpenAlex. For closed access papers, we identified PDFs by matching OpenAlex metadata using a publicly available scholarly document search API and downloaded PDFs from behind paywalls for inclusion in the analysis. We were able to download PDFs for 35K of the 60K papers. While the initial mixture of 60K papers 
% (sampled by metadata) 
were approximately 2/3 open access and 1/3 closed access, the 35K associated PDFs contained a mixture of 21.5K (61\%) open access and 13.8K (39\%) closed access papers.\footnote{This is somewhat counter-intuitive, since we assume that PDFs of open access papers would be easier to find. We believe this shift in distribution is due to several factors, such as (i) noise in the open access URLs provided by OpenAlex and (ii) open access papers not actually being made available by publishers \cite{open_access}.}
% \lucy{a bit counter-intuitive; one would expect more open access papers to have available PDFs}, reflecting the difference in accessibility between papers indexed in databases and those directly accessible through PDFs. OpenAlex indexes papers based on metadata collected from diverse sources such as institutional repositories, preprint servers, publisher websites, and author websites, where journals often label the content as open access (OA) and free-of-charge. However, despite these labels, access to full-text PDFs is often restricted due to publishers' policies or subscription requirements\cite{open_access}, resulting in fewer PDFs that are truly open access.\lucy{i don't understand this argument about why there are more closed access pdfs than percent sampled}.

% due to a variety of accessibility and availability issues, marking a 41.1\% loss from our initial collection. 
We processed these 35K PDFs through the PyPDF library to remove malformed PDFs; 30K of 35K PDFs could be opened successfully by PyPDF. Finally, we used the Adobe accessibility checker to generate accessibility reports for all remaining PDFs. Of the 30K PDFs, an additional 14.3\% could not be processed by the Adobe checker (we could not generate an accessibility report).  
% Further processing through PyPDF led to an additional decrease to 30K papers, reflecting a 24.1\% loss across the sample. Around 14.3\% of these PDFs failed to process in the Adobe checker (i.e., we could not generate an accessibility report). 
The accessibility checker most commonly fails because the PDF file is password protected or the PDF file is corrupt. In both of these cases, the PDF could be considered inaccessible to the user. We excluded these PDFs from subsequent analysis. Finally, we re-balanced our dataset by subsampling 20K PDFs with relatively equal representation across subfields, years, and open- versus closed-access papers. While preprocessing and resampling introduces bias, we emphasize that all of these steps tend to identify and remove less accessible papers, and any bias introduced is likely to lead to over-estimation rather than under-estimation of accessibility compliance.
% To re-balance this dataset, we used a statistical approach to identify outliers by calculating the mean and standard deviation of paper counts per year and venue for each sub-field. Papers deviating significantly(twice the standard deviation) from the mean count of papers within those sub-fields were considered outliers and removed from the dataset, ensuring a balanced representation of 20K papers across different sub-fields and years. 
% The accessibility checker most commonly fails because the PDF file is password protected or the PDF file is corrupt \lucy{verify this latter is true; especially since we add the PyPDF preprocessing step}. In both of these cases, the PDF is inaccessible to the user. We exclude these PDFs from subsequent analysis, and finally obtained reports for 20K papers after eliminating outliers \lucy{what's an outlier?} and re-balancing \lucy{describe how rebalancing was done} the dataset to focus on the most representative documents, ensuring an even field and year-wise distribution of papers. 

This selection process resulted in a dataset sourced from 872 unique publication venues, with representation ranging from 27 to 89 venues (average of 58) per top-level field of study, placing Art at the lower end and Computer Science at the higher end of this spectrum. Additionally, our dataset has an almost equal representation of open and closed access papers, with approximately 10K papers associated with each model of publication.
The bulk of our dataset is comprised of published papers, in contrast to preprints or other non-peer-reviewed manuscripts. Selected publication venues are reputable journals, such as \textit{The Lancet} or \textit{Neurology} for Medicine, \textit{The Astrophysical Journal} and \textit{Physical Review Letters} for Physics, and various IEEE and ACM journals for Computer Science. 
%\lucy{confirm these are in the dataset; this is from the old manuscript} --> YES
% In certain instances, the mapping between publication venues and fields of study was less straightforward; for example, the publication venue \textit{Mathematical Problems in Engineering} was categorized under Mathematics rather than Engineering in our dataset. 

We extracted metadata for each PDF document in our corpus, focusing on descriptors related to its creation process. To accomplish this, we employed the PyPDF2 library to read the metadata fields '/Creator' and '/Producer', which are populated by the software used to create each PDF. We reviewed all unique PDF creation platforms associated with more than 20 PDFs in our dataset and mapped them to standardized categories. For example, platform names containing \texttt{latex}, \texttt{pdftex}, \texttt{tex live}, \texttt{tex}, \texttt{vtex pdf}, \texttt{xetex} were mapped to the broader category "LaTeX", while names containing \texttt{microsoft}, \texttt{for word}, \texttt{word} and other variants were mapped to the Microsoft Word cluster. We realize that not all Microsoft Word versions, LaTeX distributions, or other versions of the creation platform within a cluster are equal, but this normalization allows us to generalize over these clusters. We use this metadata to analyze the associations between different PDF creation platforms and the accessibility of the resulting PDF document. 
% This helps us address the question of whether some platforms produce more accessible PDFs. 
We identify 24 total clusters, and present statistics associated with the 7 largest clusters, after which there is a steep falloff in cluster size.

\subsection{Checking PDF Accessibility}
\label{subsec:measuring_accessibility}

Consistent with prior work \citep{Lazar2017MakingTF, Ribera2019PublishingAP, Nganji2015ThePD}, we elect primarily to analyze the PDFs in our corpus using the Adobe Acrobat Pro PDF accessibility checker. 
Though this tool is proprietary and requires a paid license, it is the most comprehensive industry-standard checker available. It is also conducive to large-scale analysis, in that the checker supports the ability to process thousands of PDFs in bulk.

In addition to Adobe, we also evaluated several leading non-proprietary options, 
% to find a suitable tool for assessing PDF document accessibility, 
including AxesPDF,\footnote{\href{https://www.axes4.com/en}{https://www.axes4.com/en}} PDFBox,\footnote{\href{https://pdfbox.apache.org/}{https://pdfbox.apache.org/}} PDF Inspector,\footnote{\href{https://github.com/pdfae/PDFAInspector}{https://github.com/pdfae/PDFAInspector}} PDFix Lite,\footnote{\href{https://pdfix.net/products/pdfix-desktop-lite/}{https://pdfix.net/products/pdfix-desktop-lite/}} PAVE,\footnote{\href{https://pave-pdf.org/}{https://pave-pdf.org/}} and PAC3.\footnote{\href{https://pdfua.foundation/en/pdf-accessibility-checker-pac/}{https://pdfua.foundation/en/pdf-accessibility-checker-pac/}} 
% \lucy{Something that would be nice to include in the appendix but might not be possible (and okay to skip) due to limited time is the detailed comparison of accessibility checkers you conducted previously.} 
Our evaluation criteria focused on the comprehensiveness of accessibility checks, ease of integration into our workflow, widespread usage and recognition in the academic and/or publishing communities, and adherence to the Web Content Accessibility Guidelines (WCAG) 2.1 and PDF/Universal Accessibility (PDF/UA) standards.
We noted that AxesPDF and PDFBox did not consistently extract accessibility information from PDFs, even when we found these criteria to be met. While PDFA Inspector is used in prior studies \citep{Brady2015CreatingAP}, it only analyzes three criteria, whereas our interests extend to other essential accessibility features such as the presence of alt text on figures. PDFix Lite, despite being freely available, lacked the depth of analysis found in more comprehensive tools, especially in evaluating accessibility features like reading order and alt-text evaluation.  
PAVE, promising for its web-based accessibility checks, was impractical due to its file size limitations, lengthy processing times, and generic issue descriptions without precise location details.
PAC3 emerged as a strong contender with its detailed compliance reporting and free availability. However, its adoption is limited 
% among academics 
as there is no support available for large-scale analysis. Despite this, PAC3's in-depth reporting on standards such as WCAG and PDF/UA compliance made it an invaluable tool for secondary assessment. Hence, we utilized it to further validate our findings and address any limitations of the Adobe Acrobat Accessibility Checker.

% However, its usability challenges and the tendency for false positives in accessibility metrics, including heading structure, tab order, alternative text, and tagged content, limited its broader adoption among academics unfamiliar with PDF accessibility nuances. 

For each PDF, we used the Adobe accessibility checker to generate an accessibility report. 
% The report includes whether the PDF passes, fails, or needs manual evaluation for different accessibility criteria.
% , such as the inclusion of figure alt text or properly tagged headings for navigation. 
% Because there is no API or standalone application for the Adobe accessibility checker, it can only be 
We access the checker through the user interface of a licensed version of Adobe Acrobat Pro, and use the Action Wizard feature for bulk processing. Each PDF took 10 seconds on average to process and produce a corresponding accessibility report, allowing us to scale up our analysis to tens of thousands of papers. We saved these reports from the checker in HTML format for subsequent analysis.

Each report contains a total of 32 accessibility criteria, marked as ``Passed,'' ``Failed,'' or ``Needs manual check.''\footnote{Please see \href{https://helpx.adobe.com/acrobat/using/create-verify-pdf-accessibility.html}{https://helpx.adobe.com/acrobat/using/create-verify-pdf-accessibility. html} for a description of the accessibility report.}
We focus on the following six criteria:

\begin{itemize}[topsep=2pt,itemsep=2pt, leftmargin=12pt]
    \item Default language: The document has a specified reading language.
    \item Tagged PDF: The document is tagged to specify the correct reading order.
    \item Tab order: The document is tagged such that the tab order aligns with its structure, thereby facilitating navigation within the document.
    \item Appropriate nesting: Consistent use of heading levels or tags without skipping levels, reflecting accurate document hierarchy.
    \item Alt-text: Figures have alternate text.
    \item Table headers: Tables have headers.
\end{itemize}

\noindent 
Five of these six criteria were evaluated in prior work \citep{Lazar2017MakingTF, Wang2021ImprovingTA}. We add the \emph{Appropriate Nesting} criterion since others have emphasized the importance of structural coherence in enhancing document accessibility \cite{swiss_2023}. For these criteria, we compute and report the following:

\begin{itemize}[topsep=2pt,itemsep=2pt, leftmargin=12pt]
    \item \textit{Criteria Compliance}: for each of the 6 criteria, whether a paper passed;
    \item \textit{Total Compliance}: the sum number of accessibility criteria met  by a paper (e.g., if a paper meets 3 out of 6 criteria, Total Compliance is 3); 
    \item \textit{Normalized Total Compliance}: the proportion of the 6 criteria which are satisfied by a paper (this yields a value between 0 and 1, where 0 indicates the paper met none of the criteria and 1 means the paper met all criteria);
    \item \textit{Adobe-6 Compliance}: a binary judgment of whether a paper has met all 6 criteria (1 if all 6 criteria are met, 0 if at least one unmet)
\end{itemize}

% \noindent We report the compliance rate, representing the proportion of papers that passed each accessibility criterion.

For papers containing no tables and/or figures, we observe that the Adobe checker can still yield either a pass or fail judgment for the Table header and Alt text criteria. Specifically, when objects in the PDF are \textit{not} tagged, the checker tends to fail these criteria, even if the paper does not include tables or figures. Conversely, when objects in the PDF \textit{are} tagged and the PDF is accessible, the checker often passes these criteria, even in the absence of tables or figures. Additionally, we note that if tables are not appropriately tagged as tables (and interpreted as paragraphs instead), the Table headers criteria may also pass.

%% file: 3_sos_results.tex
\section{Analysis Results}
\label{sec:sos}

% We aim to answer the following questions with our analysis:

% \begin{itemize}
%     \item[RQ1] What is the overall status and trends in scholarly document accessibility?
%     \item[RQ2] What is the association between the software used to create a PDF and its accessibility?
%     \item[RQ3] What is the association between models of publishing (open versus closed access) and scholarly document accessibility?
% \end{itemize}

% \lucy{describe research questions earlier in intro}

\subsection{Proportion of Papers with Accessible PDFs}
\label{sec:sos_fos}

\input{figtabs/tab_compliance.tex}
\input{figtabs/fig_total_compliance.tex}
\input{figtabs/fig_compliance_fos.tex}

Accessibility compliance over all papers is low. Table~\ref{tab:compliance} shows the percent of papers meeting each of the six criteria, as well as the Adobe-6 Compliance rate associated with the sample of papers. Figure~\ref{fig:fos-total-compliance} shows that the vast majority of papers do not meet any of the six accessibility criteria (14970 papers, 74.86\% do not meet any criteria) and very few (634 papers, 3.2\%) meet all six. Of the PDFs meeting 1 criterion, the most commonly met criterion is Default Language (1750 of 1802, 97.1\%). Of the PDFs meeting 5 criteria, the most common \textit{missing} criterion is Alt-text (364 of 581, 62.6\%). In fact, only 1691 PDFs (8.5\%) in the whole dataset passed the alt-text check for figures. This is intuitive as Alt-text is the only criterion that typically requires author input to achieve, while the other five criteria can be derived from the document or automatically inferred, depending on the software used to generate the PDF. We also observed a very small number of papers (168) meeting 2 criteria (Total Compliance = 2). Among these, 3 combinations frequently occurred together (Appropriate nesting with Tagged PDF, Table Headers, or Alt Text), accounting for approximately 77\% of these papers, while combinations of Appropriate nesting with more than one of the other criteria occurred more frequently, leading to more papers that meet 3 criteria. Refer to Appendix \ref{sec:criteria_correlation} for correlations between different accessibility criteria.
% This pattern suggests that these combinations may have a stronger association and tend to occur together when other criteria are not met. However, other combinations of criteria were less common, each comprising fewer than 10 papers.

As shown in Figure~\ref{fig:fos-complete-compliance}, all fields have an Adode-6 Compliance of less than 7\%. The fields with the highest rates of compliance are Environmental Science (6.6\%), Medicine (4.9\%), Psychology (4.5\%), and Biology (4.4\%) while the fields with the lowest rates of compliance are Physics (1.8\%), Mathematics (2.1\%), and Geography (2.1\%). Fields associated with higher compliance tend to align with life sciences and health-related disciplines.
% , which may prioritize accessibility due to their broad public health implications and the diverse audience they serve, including practitioners, policymakers, and the general public, who may require accessible content.
Conversely, fields with lower compliance rates are a mixture of physical sciences and mathematical disciplines, business, and social science. Lower accessibility in math-adjacent disciplines could be attributed to the prevalence of complex equations, figures, and specialized formatting in their literature, which are challenging to make compliant without significant effort or specific knowledge of accessibility standards. Also, this variance in compliance rates across different fields may result from different field-specific strategies employed, including document editing and creation platforms, and we explore these associations in Section~\ref{sec:sos_pdf_headers}.

\subsection{Trends in Paper Accessibility Over Time}
\label{sec:trends_over_time}

\input{figtabs/fig_compliance_over_time}

We show changes in compliance for all criteria over time in Figure~\ref{fig:fos-over-time}. With the exception of Default Language, all accessibility criteria demonstrate decreasing compliance rates over the past decade, with a notable drop following 2019. Although rates of decrease varied, the relative ranking of criteria compliance remained mostly consistent; in order from most to least compliant, these are Appropriate Nesting, Table Headers, Tagged PDF, Tab Order, and Alt-Text. 
% We attribute this trend to several factors, including the rapid evolution of digital publishing technologies, which may have outpaced the integration of accessibility features. Additionally, the complexity of implementing certain accessibility features in PDF documents, such as Alt-Text for figures or ensuring proper Tab Order, may have contributed to this decline as content creators and publishers navigated the balance between advancing digital features and meeting accessibility standards. \lucy{last two sentences are more appropriate in the discussion section. results go here, speculation and explanation go in the discussion section}
% Although the rates at which these criteria, except Default Language, decreased varied, 

% However, in recent years, there have been stable or slightly increasing compliance rates across these metrics. The impact of global initiatives promoting digital accessibility and inclusivity, alongside advancements in typesetting software, might have contributed to these improvements.

Compliance with the Default Language criterion has seen the most rapid increase, from around 10\% in 2014 to over 25\% in 2023. This may be due to changes in PDF generation defaults in various creation platforms, making it easier for authors or publishers to comply with this particular criterion. Though this improvement is positive, it is important to note that ensuring a default language is set is the easiest of the six criteria to bring into compliance, and arguably the least valuable in terms of improving the accessible reading experience. 
The criteria showing the lowest rates of compliance are Tab Order and Alt-Text. These findings are particularly concerning because Alt-Text is the only criterion among the six that requires direct input from authors, 
% Given our previous observation that most Alt-Text among papers passing the accessibility checker lacked sufficient information about the associated figure, 
suggesting a continued lack of engagement with accessibility considerations among authors. It also further supports the theory that any observed improvements can be partially attributed to automatic enhancements made by creation platforms or publisher-level changes rather than a genuine increase in accessibility awareness and compliance among content creators.
Furthermore, the general decline in compliance across most criteria, except for Default Language, also resulted in a downward trend in Adobe-6 Compliance. This trend reinforces the notion that despite technological advancements and potential shifts in publisher policies, significant gaps in accessibility remain, underscoring the need for continued efforts to raise awareness and improve compliance with all accessibility criteria, not just the simplest ones.

\subsection{Association Between PDF Creation Platform and Paper Accessibility}
\label{sec:sos_pdf_headers}
\input{figtabs/fig_compliance_typesetting.tex}
For analysis, we compare the seven most commonly observed creation platform clusters in our dataset, grouping all others into a cluster called \texttt{Other}. The most popular PDF creators are Arbortext APP, Adobe InDesign, LaTeX, Microsoft Word, XPP, Springer, and Elsevier. Among these, Arbortext APP is used predominantly for technical documentation due to its precision in layout control, while Adobe InDesign is preferred for its advanced layout capabilities across both digital and print media. LaTeX is highly esteemed in academia for its ability to handle complex mathematical formulas and structures effectively. Microsoft Word is widely utilized for its versatility in document editing and formatting across various settings. XPP, or Xyvision Production Publisher, provides advanced automation in creating complex documents, essential in professional publishing environments. Springer and Elsevier are not software but well-known publishing companies that often employ their own proprietary tools or standardized templates to create and typeset documents according to their specific publishing standards. The ``Other'' category aggregates papers created by the remaining 17 clusters of creation platforms, each with counts of less than 30, as well as those created with unknown PDF creation platforms. 
% For the following analysis, we present a comparison between the five most common PDF creator clusters.

% \input{figtabs/tab_typesetting.tex}
Figure~\ref{fig:fos-total-compliance-headers} shows histograms of the Total Compliance score for PDFs in the seven most common PDF creation platform clusters. While the vast majority of papers do not meet any accessibility criteria, it is clear that Microsoft Word produces the most accessible PDFs, followed by Elsevier, Adobe InDesign, and Springer. 
To determine the significance of this difference, we apply the
Kruskal-Wallis $H$-test \citep{Kruskal1952UseOR}, a non-parametric method for analysis of variance that can be applied to non-normally distributed data. With the PDF creation platform clusters as the sample groups and the Total Compliance as the measurements for the groups, we compute 
a Kruskal-Wallis $H$ statistic of 6378.8 ($p$ < 0.001). This indicates a significant difference in the distribution of Total Compliance scores between the seven most common PDF creation platforms. 
% Microsoft Word in particular, demonstrates significantly higher accessibility compliance than other typesetting software.

% \input{figtabs/fig_word_by_fos.tex}
% \input{figtabs/fig_prop_software_compliance}
% \input{figtabs/fig_latex_by_fos.tex}
% \input{figtabs/fig_compliance_software_time.tex}

In Appendix ~\ref{sec:regression_analysis}, we provide plots of correlations between the proportion of PDFs typeset using different creation platforms per field of study and the corresponding mean normalized Total Compliance rates for those fields. Higher rates of Microsoft Word usage are statistically correlated with higher mean normalized Total Compliance ($r = 0.66$, $p < 0.01$). Conversely, higher rates of LaTeX usage are statistically correlated with lower mean normalized Total Compliance ($r = -0.74$, $p < 0.001$). There is also a positive correlation between PDFs created with header Elsevier and higher accessibility compliance ($r = 0.53$, $p < 0.05$). Correlation scores for Adobe InDesign, Springer, and XPP are non-significant (r = 0.16, 0.12, and -0.06 respectively, all non-significant).

% Microsoft Word and LaTeX usage per field of study and the corresponding mean normalized Total Compliance rates for those fields. Higher rates of Microsoft Word usage are statistically correlated with higher mean normalized Total Compliance ($r = 0.66$, $p < 0.01$). Conversely, higher rates of LaTeX usage are statistically correlated with lower mean normalized Total Compliance ($r = -0.74$, $p < 0.001$). Correlations for other large creation platform clusters are included in Appendix \ref{sec:regression_analysis}.
% \lucy{add regressions for other software to appendix; or possibly add all as subplots}

% \input{figtabs/fig_typesetting_over_time.tex}

\input{figtabs/fig_typeseting_over_time}
We plot changes in the usage of each of the seven PDF creation platforms over time in Figure~\ref{fig:software_over_time}. Early in the decade, papers were predominantly created using Arbortext APP and Adobe InDesign. Recently, however, there has been a noticeable growth in the usage of LaTeX, Microsoft Word, and Springer. Conversely, Arbortext APP, Elsevier, and XPP are witnessing a gradual decline in usage, suggesting they are being replaced in PDF headers, becoming obsolete, or are less preferred compared to their alternatives. These shifts in preferences reflect broader changes in the publishing industry, driven by evolving standards and technological advancements. Such changes are often influenced by endorsements from leading institutions and journals, contributing to significant trends in document accessibility. Refer to Appendix ~\ref{sec:typesetting_software_over_time} to examine trends of creation platforms used in open versus closed publishing models over time. Splitting by open versus closed access shows large differences, with platforms like Adobe InDesign and LaTeX favored among open access papers along with the rise of Microsoft Word, while platforms like Arbortext APP were heavily used in the earlier part of the decade among closed access papers and dominate yet along with LaTeX.

\input{figtabs/fig_compliance_software_time}
% To further assess the influence of major PDF creation platforms (Microsoft Word, Adobe InDesign, and LaTeX) on accessibility compliance over time, we plot the mean normalized compliance of papers created using these three platforms over time in Figure ~\ref{fig:compliance_software_time} \lucy{this can go in appendix too. keep 2-3 sentences of discussion about PDF creation platform changes over time and association with accessibility compliance here in the main text}. While accessibility compliance for PDFs created with LaTeX remains relatively stable (and inaccessible), a decline in compliance is observed for PDFs created using Microsoft Word and Adobe InDesign, especially for the period of time from 2017 to 2020. This trend aligns with patterns in overall compliance scores over time, as shown in Figure~\ref{fig:fos-over-time}. 
To further assess the influence of major PDF creation platforms (Microsoft Word, Adobe InDesign, and LaTeX) on accessibility compliance over time, we plot the mean normalized compliance of papers created using these three platforms over time in Figure ~\ref{fig:compliance_software_time}. While accessibility compliance for PDFs created with LaTeX remains relatively stable (and inaccessible), a decline in compliance is observed for PDFs created using Microsoft Word and Adobe InDesign, especially for the period of time from 2017 to 2020. This decline may be partially attributed to updates in these platforms that prioritized visual and functional enhancements without adequate integration of accessibility standards. For example, the 2019 updates to Microsoft Office introduced AI-driven tools,\footnote{See \href{https://www.microsoft.com/en-us/microsoft-365/blog/2019/06/18/powerpoint-ai-upgrade-designer-major-milestone-1-billion-slides/}{https://www.microsoft.com/en-us/microsoft/blog/2019/powerpoint-ai-upgrade-designer} and \href{https://venturebeat.com/ai/6-ai-features-microsoft-added-to-office-in-2019/}{https://venturebeat.com/ai/features-office-2019/}} designed to improve user productivity did not include necessary accessibility features like tagging or alt text. Furthermore, the adoption of these new features without corresponding increases in accessibility awareness or training likely contributed to the observed drop in compliance. This trend aligns with the overall patterns in compliance scores over time, as detailed in Figure~\ref{fig:fos-over-time}.

\subsection{Association Between Models of Publishing and Paper Accessibility}
\input{figtabs/fig_compliance_over_models}
Our dataset contains near equal representation of open (10.1K) and closed (9.9K) access models of publishing.
% , namely open and closed access papers in our dataset, accounting for 10.1K and 9.9K papers respectively. 
Our hypothesis was that open access papers would demonstrate higher accessibility compliance compared to closed access papers due to the broader visibility and inherent motivation for wider dissemination in the open access model. To assess this hypothesis, we studied accessibility compliance across these two access types, as illustrated in Figure~\ref{fig:compliance-over-models}. The median of accessibility compliance over fields of study for open-access papers was 0.18 (IQR = 0.04), significantly higher than the 0.07 (IQR = 0.01) observed for closed-access papers. 

To determine the significance of this difference, we applied the Mann-Whitney U test, a non-parametric method suitable for comparing the distributions of two independent groups with non-normally distributed data. With the publishing access type (open vs. closed) as the sample groups and the accessibility compliance as the measurements for these groups, we found a significant difference in the distribution of mean normalized total compliance scores between the two access types (Z = 24.76, p < 0.0001). 

Figure~\ref{fig:compliance_model_time_errors} shows compliance changes by publication model (open vs. closed access) 
% across all fields of study 
over time. The plot reveals a large decline in mean normalized total compliance for open access papers after 2019, followed by relatively stable or slightly increasing compliance values. These trends suggest that the decline in accessibility compliance observed in Figure~\ref{fig:fos-over-time} can largely be attributed to a decline in compliance among open access publications. 
% This trend mirrors the pattern observed in overall compliance scores over time in Figure~\ref{fig:fos-over-time}. 
This finding supports our initial hypothesis and highlights that publishing model has a strong influence on PDF accessibility compliance. While improvements in accessibility compliance are still needed across the board, the emerging crisis among open access publications begs further attention and marks an unwelcome departure from observed improvements in scholarly PDF accessibility documented in prior work.

%authors indeed make extra efforts to enhance the accessibility of academic papers when publishing open access.

% We have an almost equal representation of the models of publishing, namely open and closed access papers in our dataset, accounting for 10.1K and 9.9K papers respectively. Our hypothesis was <>. To assess this hypothesis, we study accessibility compliance over these two access types in Figure 8. The median of mean normalized total compliance over all fields of study for open-access papers and closed-access papers were 0.18 (IQR = 0.04) and 0.07 (IQR = 0.01). To determine the significance of this difference, we apply the Mann-Whitney U test [], a non-parametric method for analysis of variance that can be applied to non-normally
% distributed data with 2 groups. With the PDF access type clusters as the sample groups and the Mean Normalized Total Compliance as the measurements for the groups, we find a significant
% difference in the distribution of Mean Normalized Total Compliance scores between the two access types (Z = 24.76, p < 0.0001). Open access papers, in particular, demonstrates significantly higher accessibility compliance than the closed-access ones.

\section{Assessing the Limitations of Automated Accessibility Checkers}
\label{sec:sos_evaluate_checkers}
While we rely on automated accessibility checkers to enable large-scale analysis, we acknowledge they are imperfect instruments. We therefore conduct additional analysis to qualify the limitations of these methods, through manual evaluation of alt text, comparative assessments with an alternate accessibility checker, and manual assessments with screen readers.

\subsection{Verifying alt-text quality}
\label{sec:sos_alttext_quality}

% \lucy{I'm having a hard time figuring out where this subsection should go; since it's analysis conducted on the whole dataset, one could argue it should go before the PAC3 and screen reader assessments. Moved it up here for now to match methods}
We further process PDFs that ``Passed'' the alt-text criteria and extract the corresponding author-written alt-texts. To extract alt-text, we employ the method described in \citet{Chintalapati_2022}, which uses the Adobe Acrobat Pro PDF to HTML conversion utility to convert the sample of documents into HTML, from which we access the alt-text associated with each figure. We successfully converted and extracted alt-texts from 1536 of 1691 PDFs that “Passed” the Alt-text criteria. 
Given that significant information content is required of figure alt-text to satisfy BLV user needs, we analyzed these alt-texts to determine whether they contain meaningful descriptions of figure content. Although we initially intended to use frameworks from prior work \citep{Lundgard2022AccessibleVV, Mack2021DesigningTF, Williams} to assess alt-text quality, we found that many alt-texts extracted from our data sample were at or below the lowest evaluative levels of these frameworks. As a result, we manually categorized them into five types: (i) a filepath, (ii) unspecific (words like ‘Image’ or ‘Figure’ that are likely auto-generated), (iii) caption (duplicate of the figure caption), (iv) cursory (too short to provide a meaningful understanding of the image), and (v) satisfactory (contains information about the visual aspects of the image).

% Given that significant information content is required of figure alt-text to satisfy BLV user needs \citep{Lundgard2022AccessibleVV, Mack2021DesigningTF, Williams}, we analyzed these alt-texts to determine whether they contain meaningful descriptions of figure content. We manually categorized them into five types: (i) a filepath, (ii) unspecific (words like ‘Image’ or ‘Figure’ that are likely auto-generated), (iii) caption (duplicate of the figure caption), (iv) cursory (too short to provide a meaningful understanding of the image), and (v) satisfactory (contains information about the visual aspects of the image). 

On analyzing these, we found that 95\% of these papers (1461) contained nondescript alt-text such as ‘Image,’ ‘Figure,’ or ‘Logo.’ Approximately 3.2\% (49) papers had the file path or URL of the image source in their alt-texts, and 1.4\% (21) had minimal descriptions such as `A picture containing graphical user interface', `A close-up of a map', and `A picture containing surface chart', which may not offer sufficient information to understand the visual content of the image. Only 5 PDFs (less than 0.4\%) that passed the Alt-text criterion (already only 8.5\% among our total sample of 19,997 PDFs) actually contained useful and meaningful alt-text, offering detailed descriptions of figure content. Extrapolating to our entire sample, this equates to around 0.03\% of papers meeting the stricter criteria of having “meaningful” alt-text. 
% The Adobe checker evaluates alt-text compliance by verifying the presence of alt-text but does not assess their descriptiveness. We extracted and labeled alt-text from all PDFs in our sample that passed the alt-text criteria, categorizing these alt-texts into five types (filepath, unspecific, caption, cursory and satisfactory), as described in Section \ref{subsec:extracting_alttext}. Our findings indicate that 95\% of these papers (1461) contained nondescript alt-text such as ‘Image,’ ‘Figure,’ or ‘Logo.’ Approximately 3.2\% (49) papers had the file path or URL of the image source in their alt-texts, and 1.4\% (21) had minimal descriptions such as `A picture containing graphical user interface', `A close-up of a map', and `A picture containing surface chart', which may not offer sufficient information to understand the visual content of the image.
% auto-generated by typesetting software during PDF creation. 
% Only 5 PDFs (less than 0.4\%) that passed the Alt-text criterion (already only 8.5\% among our total sample of 19,997 PDFs) actually contained useful and meaningful alt-text, offering detailed descriptions of figure content. Extrapolating to our entire sample, this equates to around 0.03\% of papers meeting the stricter criteria of having “meaningful” alt-text. 

% \lucy{this an next section might belong in the discussion; or the checker evaluation in its own section}

\subsection{Comparative assessment with Adobe Checker and PAC3}
\label{sec:compare_pac3}
\input{figtabs/tab_checker_comparison}
% In our study, 
We extend the evaluation of PDF accessibility beyond Adobe’s checker by incorporating a secondary assessment with the PAC3 tool. Because the tool has no API or bulk processing feature, assessment is manual and we are forced to limit our analysis to a smaller sample of PDFs. We subsample 133 papers, stratified along the 19 top-level fields of study across total compliance values 0-6 as determined by the Adobe checker. For each of these PDFs, we manually conduct a check using PAC3 and save the report for analysis. Unlike Adobe, which reports element-wise compliance, PAC3 reports are organized based on the WCAG and PDF/UA standards. The criteria reported by these checkers are not equivalent, so we manually identify correspondences to our six assessed criteria in the PAC3 output (Table~\ref{tab:checker_comparison}) and extract these values for analysis.

We compare the outputs of these two checkers: Adobe and PAC3, side by side along with the original PDF. Our primary tool for this analysis was Adobe Acrobat Reader, supplemented by manual testing with screen readers (VoiceOver), especially to assess the practical navigational experience with the keyboard. For the Tagged PDF, Table Headers, and Appropriate Nesting criteria, we utilized the Reader to inspect the presence and hierarchy of PDF tags within the tag tree. We further used it to check for the presence of Alt-text for images. We also verified the specification of the Default Language by examining the document properties. Additionally, we evaluated the Tab Order criterion using primarily VoiceOver to verify if the tab order paralleled the document structure. These results are presented in Table~\ref{tab:checker_comparison}.\footnote{The false positive and false negative rates reported in Table~\ref{tab:checker_comparison} were measured on 152 papers, which up-samples compliant PDFs, stratified by Adobe compliance score. These rates are not representative of false positive and false negative rates over our entire corpus though serve to illustrate commonly occurring limitations of these tools.}

Our comparative analysis of the Adobe and PAC3 checkers revealed discrepancies in compliance values for specific criteria, highlighting potential disagreements between the checkers and their impact on the conclusions drawn from automated assessments. For example, our analysis reveals that the Adobe Checker is more likely to yield false positive classifications for the Table Headers criterion (saying this criterion is not met) compared to PAC3.
% For example, our analysis reveals that the Adobe Checker is more likely to yield false negative classifications for the Tagged PDF criterion (incorrectly asserting that this criterion is met), with it passing the Tagged PDF criterion in 91 PDFs (59.8\%) compared to only 29 PDFs (19\%) by PAC3. 
Conversely, PAC3 appears more lenient, often overestimating compliance, especially for the Appropriate Nesting and Table Headers criteria. For instance, PAC3 passed the majority of documents (more than 80\%) in these criteria, overlooking significant accessibility issues. This could lead to a false sense of progress regarding the state of document accessibility. These findings underscore the limitations of relying solely on automated tools for PDF accessibility evaluation and emphasize the necessity for manual reviews to accurately evaluate PDF accessibility.

\subsection{Manual assessments with screen readers}
\label{sec:manual_screen_reader}
% Our study reveals significant gaps in how automated tools evaluate the nuanced aspects of PDF accessibility. These 
Automated accessibility checker tools provide a binary `Passed' or `Failed' judgement while evaluating documents against a predefined set of standards such as WCAG and PDF/UA, but often fall short in assessing a document's practical and functional coherence. We explore the limitations of automated checkers across some of these criteria by conducting manual assessments with screen readers. We read 133 PDFs (sampled across fields and total compliance rates) using NVDA and VoiceOver, and describe key differences between checker results and the qualitative experience of using screen readers to read these documents:

\begin{enumerate}[topsep=2pt,itemsep=2pt,leftmargin=14pt]
    \item Tagged PDF: Although automated tools can verify the presence of tags in a PDF, they do not assess the correctness and appropriateness of these tags (e.g., whether a tag used as a list actually corresponds to list items), which often requires human judgment. These tools also fail to interpret the reading sequence or identify issues caused by visually overlaid content, structured differently in the tag tree. This can result in documents where the tagged order conflicts with intuitive visual cues, making navigation difficult for screen reader users.

    \item Table Headers and Appropriate nesting: Automated tools can check for the presence of table header tags, but they cannot determine if these headers are meaningful or if they are correctly associated with the corresponding data cells through attributes like scope or headers. Additionally, these tools often fail with complex tables, such as those with multi-level headers or irregular header arrangements, and thus, require manual verification. Likewise, automated checkers can flag improper heading hierarchies and validate against syntactic rules but fail to assess whether the nesting reflects the intended logical structure of the content. As a result, users might find it challenging to understand these relationships and context within the document.

    \item Tab order: While automated tools can check if tab order is specified, they cannot determine if the order makes logical sense. This misalignment often results in a disjointed reading experience, making it challenging for users to navigate the content efficiently or correctly.

    \item Default Language: These checkers cannot assess whether the detected language accurately reflects the document's content. For instance, a page could be marked as English while containing primarily French text. They are also not capable of detecting changes in language in a multilingual context.
    % , requiring screen readers to update language changes. 
    This oversight can lead to scenarios where even if a document is technically compliant, it can be practically misleading for users when the default language does not match the content.

    \item Alt-text: While these checkers are capable of detecting the presence of alternative text for images and other non-text elements, they fail to assess the descriptiveness and contextual relevance of this text. For example, an alt-text that says "image" or "logo" does not convey meaningful information about the image's content, rendering them inaccessible to screen reader users. 
    Our manual review of alt-texts in Section~\ref{sec:sos_alttext_quality} reveals that although Adobe's checker identified them as compliant, only a small fraction were genuinely descriptive and meaningful.

\end{enumerate}

% \hfill \break
Along all criteria, automated checkers fail to capture readability issues that occur when reading these documents with screen readers. 
% Our analysis underscores the necessity for manual, user-centered evaluations to complement automated assessments in ensuring PDF accessibility. 
Automated checkers, while useful for initial broad assessments, are not a substitute for human evaluation, and are limited in interpreting the quality and practicality of compliance. We ask the reader to bear this in mind when interpreting our analysis (in that we believe the actual screen reader accessibility of scholarly documents to be even lower than what our analysis reports), or when considering the use of automated accessibility checkers as the sole instrument for validating accessibility. 
To truly support the accessibility needs of all users, especially those relying on screen readers, a more comprehensive approach that combines both automated and manual review techniques is required. This will help uncover and address disparities between technical compliance and practical accessibility, ensuring more inclusive access to digital content.

%% file: figtabs/tab_compliance.tex
\begin{table}[t!]
\begin{tabular}{lc}
    \toprule
    \textbf{Criterion} & \textbf{Percentage of papers (n=19997)} \\ 
    \midrule    
    Default language & 17.3\% \\
    Tagged PDF & 12.6\% \\
    Tab order & 6.8\% \\
    Appropriate Nesting & 15.9\% \\
    Alt-text & 8.5\%  \\
    Table headers & 13.4\% \\
    \midrule
    Adobe-6 Compliance & 3.2\% \\
    \bottomrule \\ [-2mm]
\end{tabular}
\caption{Percent of papers in our dataset of 19997 PDFs that satisfy each criterion, along with Adobe-6 Compliance.
}
\label{tab:compliance}
% \Description{
% Criterion; CHI 2010 [23]; Ours-CHI 2010; Ours-All (11,397)
% Alt-text; 3.6%; 4.0%; 7.5%
% Table headers; 0.7%; 1.0%; 13.3%
% Tagged PDF; 6.3%; 7.4%; 13.4%
% Default language; 2.3%; 3.0%; 17.2%
% Tab order; 0.3%; 1.0%; 9.3%
% Adobe-5 Compliance; -; -; 2.4
% }
\end{table}

%% file: figtabs/fig_total_compliance.tex
\begin{figure}[tb!]
  \centering
    \includegraphics[width=0.95\linewidth,trim={0mm 8mm 0mm 0mm},clip]{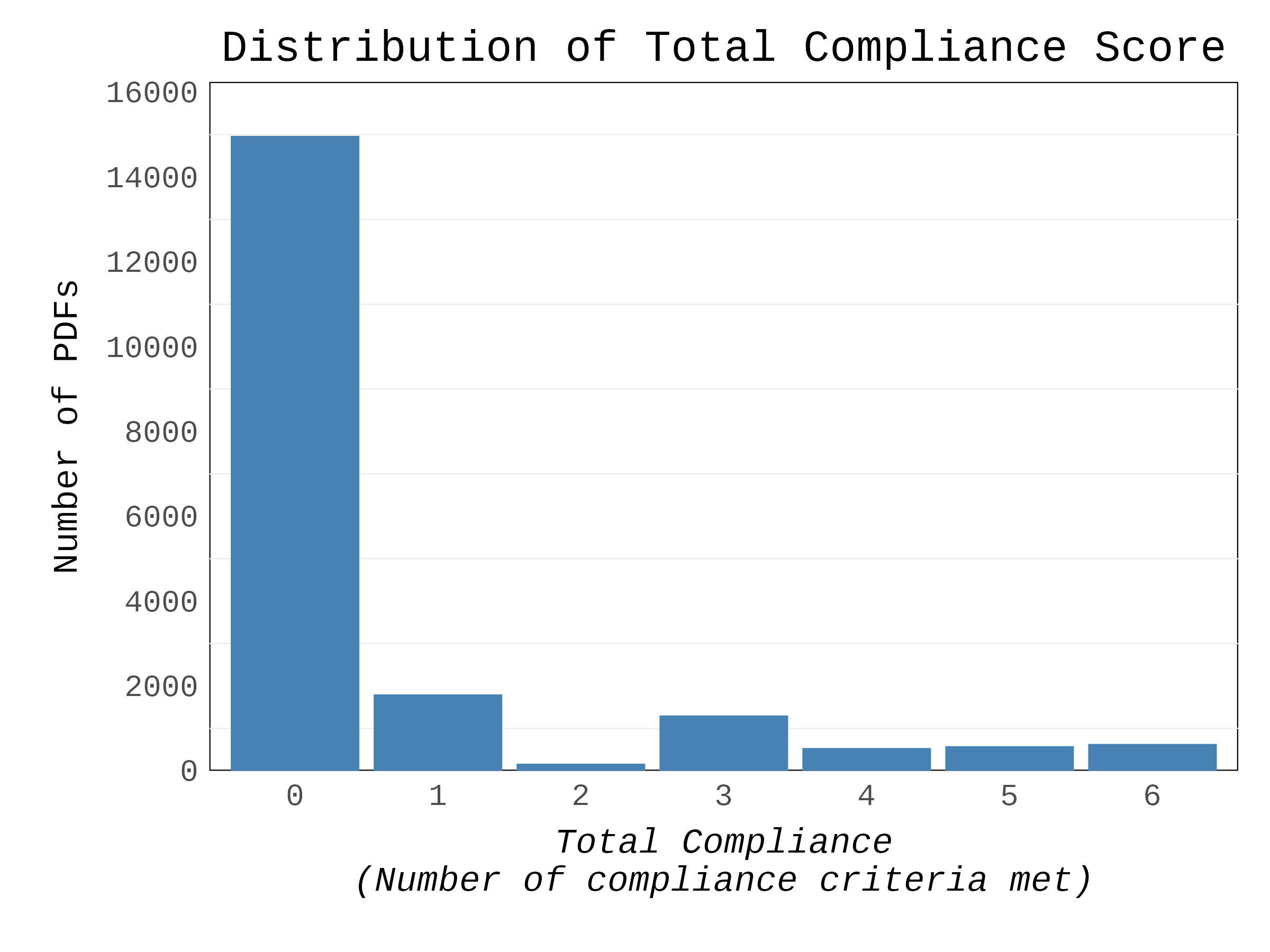}
  \caption{The distribution of numbers of PDFs in our dataset that meet our defined accessibility compliance criteria. A large majority (14970, 74.9\%) meet 0 out of 6 accessibility criteria. Of those meeting 1 criterion (Total Compliance = 1), the most commonly met criterion is Default Language (1750 of 1802, 97.1\%). Of those meeting 5 criteria (Total Compliance = 5), the most common missing criterion is Alt-text (384 of 581, 66.1\%). A small number of PDFs (168) meet only 2 criteria (Total Compliance = 2). Among these, 3 combinations frequently occurred together (Appropriate nesting with Tagged PDF, Table Headers, or Alt Text), accounting for approximately 77\% of the papers. This suggests a stronger association among these combinations.
  } 
  \label{fig:fos-total-compliance}
  \Description{A histogram shows the distribution of total compliance scores within our dataset. The x-axis indicates the total compliance, ranging from 0 to 6, and the y-axis shows the count of PDFs with corresponding compliance levels. Most PDFs (14,970 out of 19,997) do not meet any compliance criteria. A small subset of PDFs, 168 in total, meet exactly 2 criteria. Small numbers of PDFs meet some criteria, with lower numbers meeting more criteria, with less than 3.2\% of PDFs meeting all the criteria.} 
\end{figure}

% \lucy{check why compliance 2 has no papers; add a comment to explain}

%% file: figtabs/fig_compliance_fos.tex
\begin{figure}[t!]
  \centering
    \includegraphics[width=\linewidth,trim={5mm 7mm 3mm 2mm},clip]{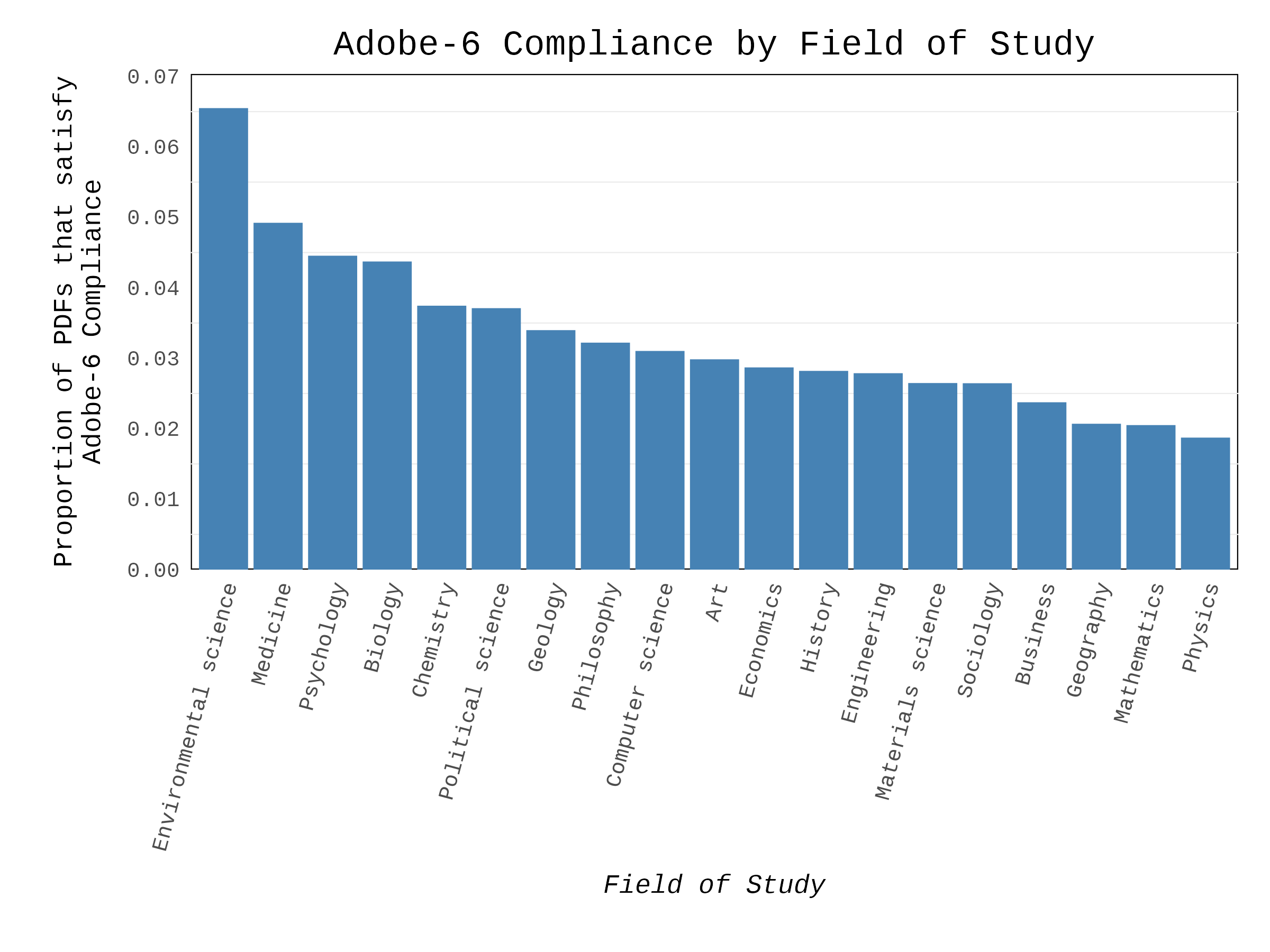}
  \caption{Proportion of papers per field of study that meet all six accessibility criteria defined by Adobe-6 Compliance. Environmental Science (6.6\%), Medicine (4.9\%), Psychology (4.5\%), and Biology (4.4\%) have the highest rates of Adobe-6 Compliance while the fields
with the lowest rates of compliance are Physics (1.8\%), Mathematics (2.1\%), and Geography (2.1\%). None of these fields had more than 6.6\% of PDFs satisfying all the six criteria.
  }
  \label{fig:fos-complete-compliance}
  \Description{A bar plot shows the proportion of PDFs in each field of study that satisfy Adobe-6 Compliance (meets all six accessibility criteria we define). Compliance percentage ranges from 6.6\% at the high end to 1.8\% at the low end. At the high end are fields such as Environmental Science, Medicine, Psychology, and Biology. At the low end are fields like Physics, Mathematics, and Geography.}
\end{figure}

%% file: figtabs/fig_compliance_over_time.tex
\begin{figure}[t!]
  \centering
    \includegraphics[width=\linewidth]{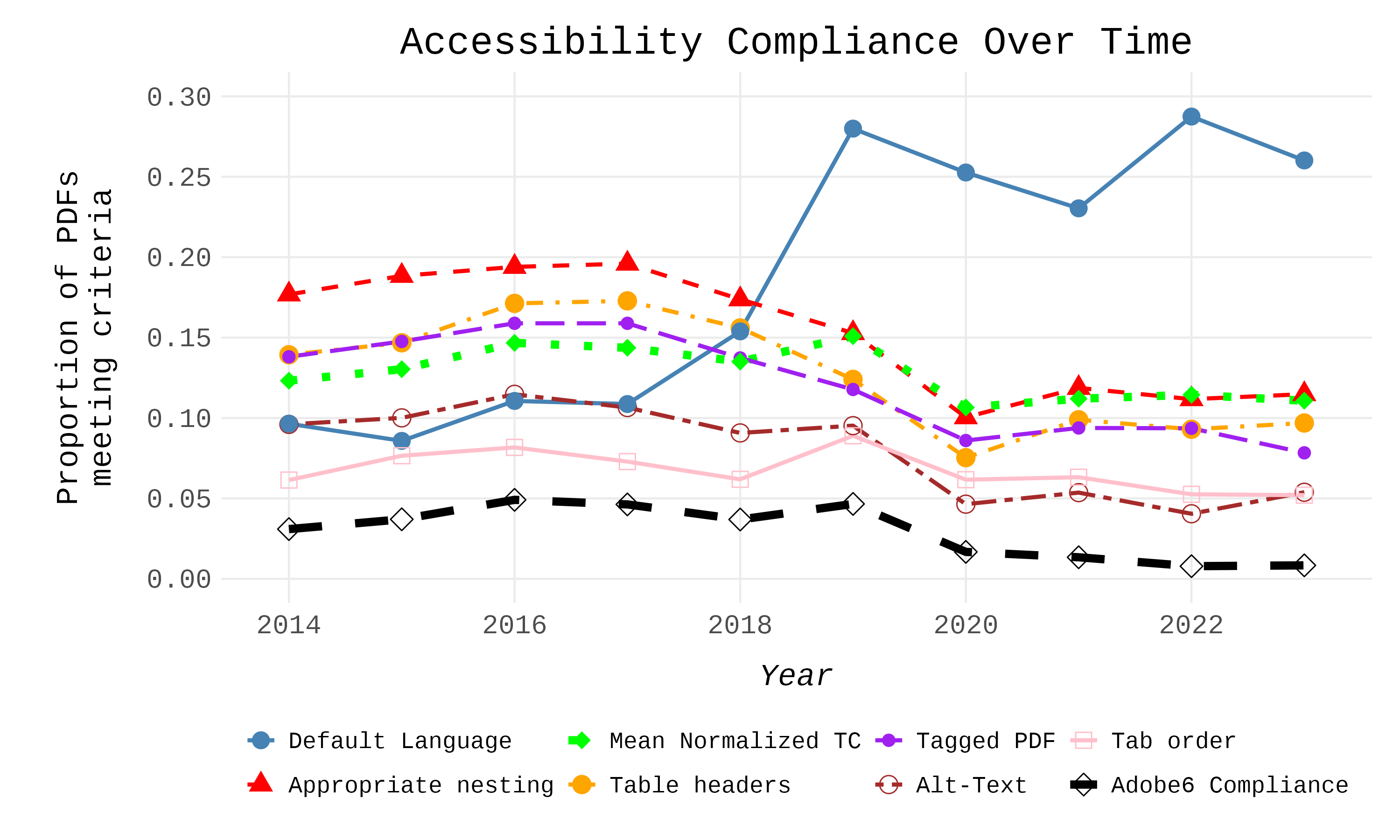}
  \caption{Accessibility compliance over time (2014-2023). While compliance rates seemed to improve slightly over time up through 2017, we observe a large decrease in all criteria but Default Language since then. There is also a significant decline in the mean normalized total compliance and Adobe-6 compliance after 2019. This is a concerning trend in the wrong direction that we attribute to a lack of improvement in accessibility awareness as well as changes in PDF creation platforms and models of publishing. 
  % \lucy{would be good to look this plot for *only* microsoft word papers or *only* latex papers to see if the trend is dominated by changes in typesetting software used}
\\  }
  
  \label{fig:fos-over-time}
  \Description{A line plot shows changes in compliance rates from 2014 to 2023 across specific criteria (Appropriate nesting, Tagged PDF, Table Headers, Default Language, Alt-Text, and Tab order), along with the mean normalized total compliance and Adobe-6 compliance. The x-axis denotes the years, and the y-axis shows the proportion of PDFs in our sample that meet the criteria. Notably, since 2017, there has been a significant decline in the proportion of PDFs compliant with all criteria except for Default Language. After 2019, both the mean normalized total compliance and Adobe-6 compliance also show a decline. In contrast, the proportion of PDFs meeting the Default Language criterion has notably increased from 0.10 in 2014 to 0.27 in 2023.}
\end{figure}

%% file: figtabs/fig_compliance_typesetting.tex
\begin{figure}[t!]
  \centering
    \includegraphics[width=\linewidth]{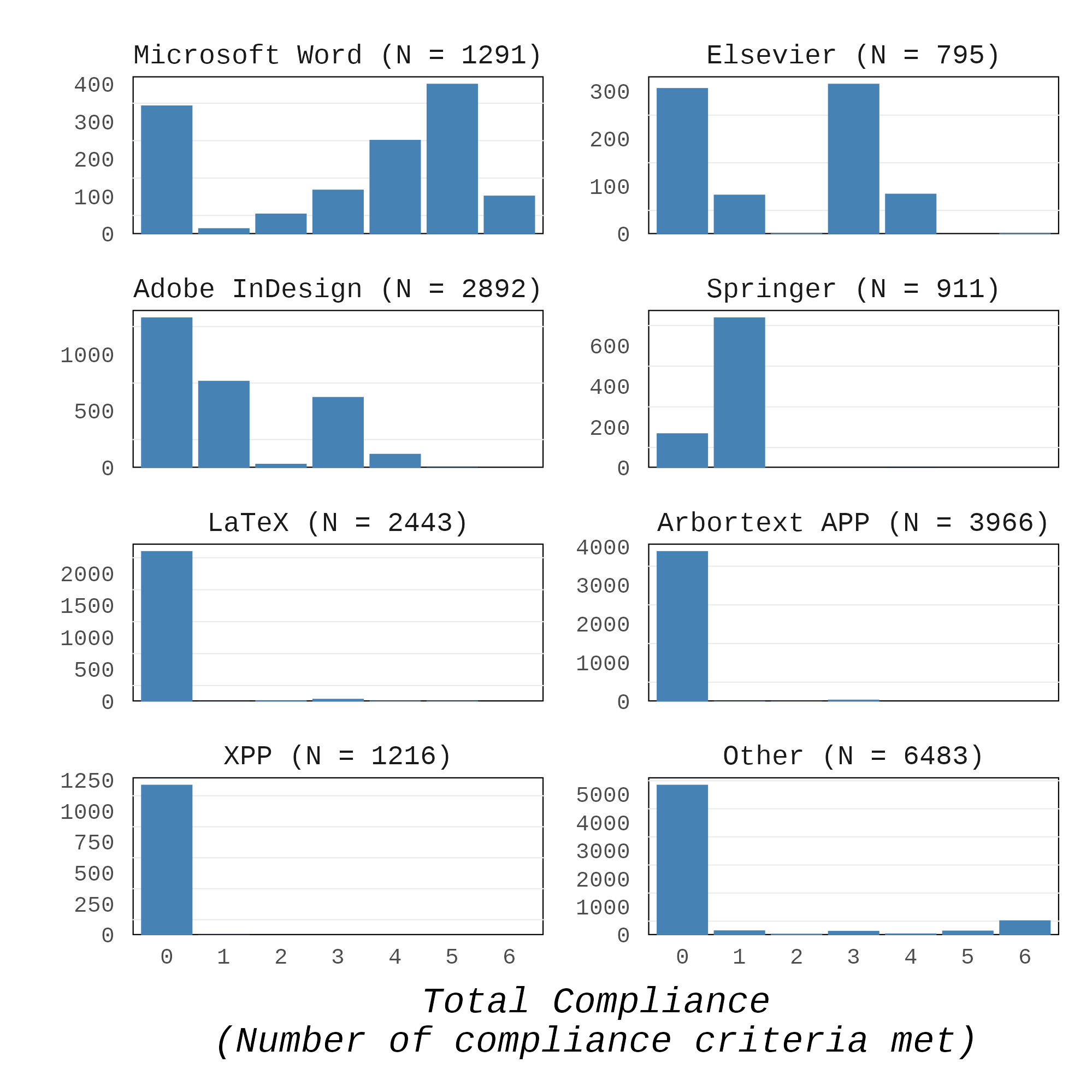}
  \caption{Histograms showing the distribution of Total Compliance scores for each of the top 7 PDF creation platforms, and an 'Other' category that groups the remaining platforms, ordered by decreasing mean Total Compliance. Microsoft Word stands out as producing PDFs with significantly higher Total Compliance than other platforms. Three of the most common PDF creation platform clusters, Arbortext APP, XPP, and LaTeX, produce PDFs with low Total Compliance, with the majority of PDFs at 0 compliance.
  }
  \Description{Eight histograms show the distribution of Total Compliance scores for the seven most common PDF creation platform clusters, and the distribution of score for all Other creation platforms. These are sorted from most compliant to least compliant, in order: Microsoft Word (n=1291), Elsevier (n=795), Adobe InDesign (n=2892), Springer (n=911), LaTeX (n=2443), Arbortext APP (n=3966), and XPP (n=1216). The Other category includes n=6483 PDFs. Microsoft Word produces many PDFs that satisfy 2 or more criteria, with a peak at Total Compliance = 5. Elsevier produces PDFs most commonly satisfying 3 criteria, followed by 0 criteria. Most PDFs produced by Adobe InDesign satisfy no accessibility criteria, but many satisfy 1 or 3. Springer produces PDF typically satisfying 1 criteria. Arbortext APP, LaTeX, and XPP all produce inaccessible PDFs, with the vast majority of PDFs produced by these software satisfying no accessibility criteria.}
  \label{fig:fos-total-compliance-headers}
\end{figure}

%% file: figtabs/fig_typeseting_over_time.tex
\begin{figure}[h!]
  \centering
    \includegraphics[width=\linewidth]{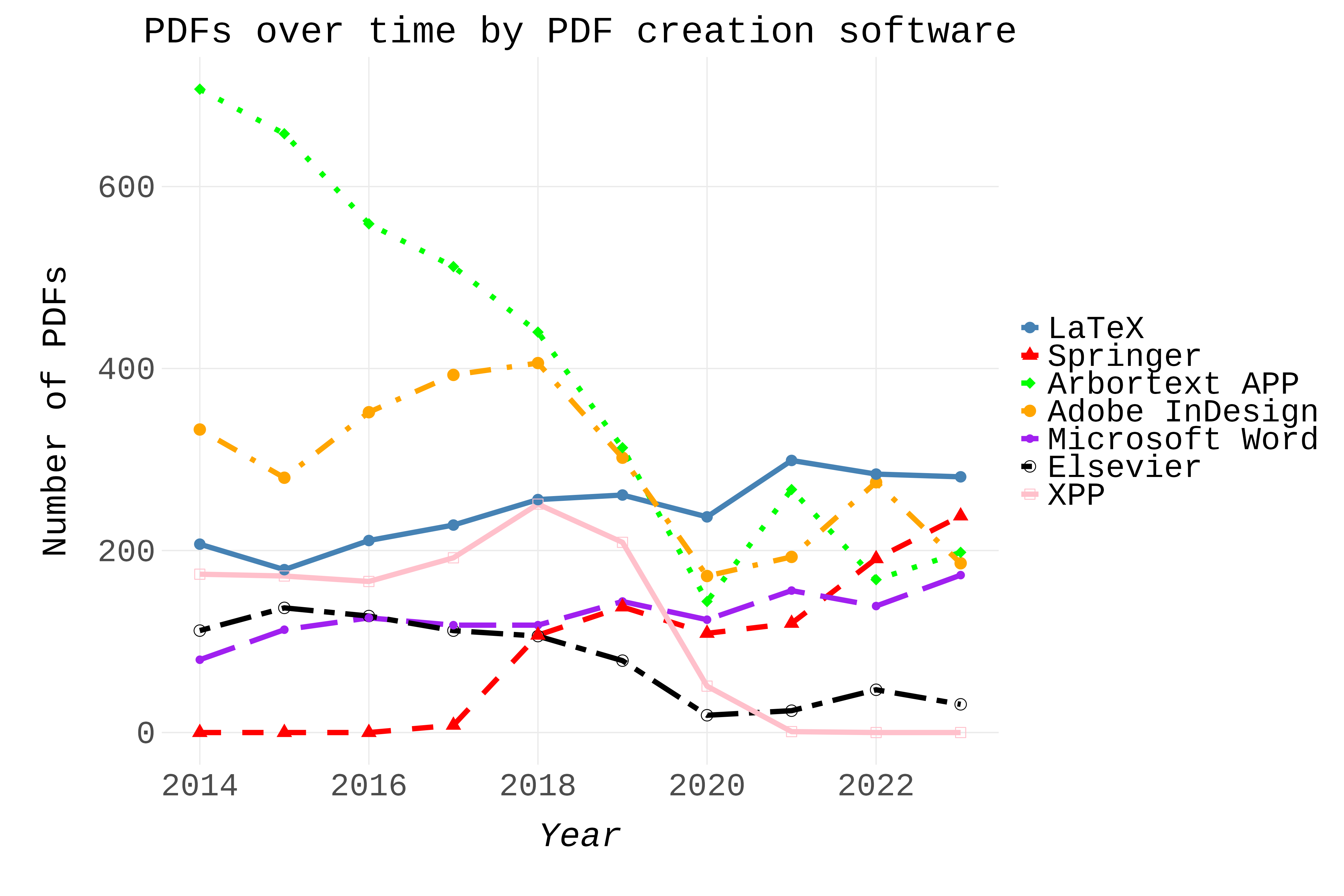}
  \caption{The number of PDFs created by the seven most common PDF-creating platforms over time. Overall, platforms such as LaTeX, Microsoft Word, and Springer are increasing in popularity over time. Conversely, the popularity of Arbortext APP, Adobe InDesign, Elsevier, and XPP has declined significantly.}
  \label{fig:software_over_time}
  \Description{A line plot shows the number of PDFs created using different creation platforms in our sample from 2014 to 2023. It shows a rising trend in the use of LaTeX, Microsoft Word, and Springer. In contrast, the popularity of Arbortext APP, Adobe InDesign, Elsevier, and XPP has declined significantly over the decade.}
\end{figure}

%% file: figtabs/fig_compliance_software_time.tex
\begin{figure}[t!]
  \centering
    \includegraphics[width=\linewidth]{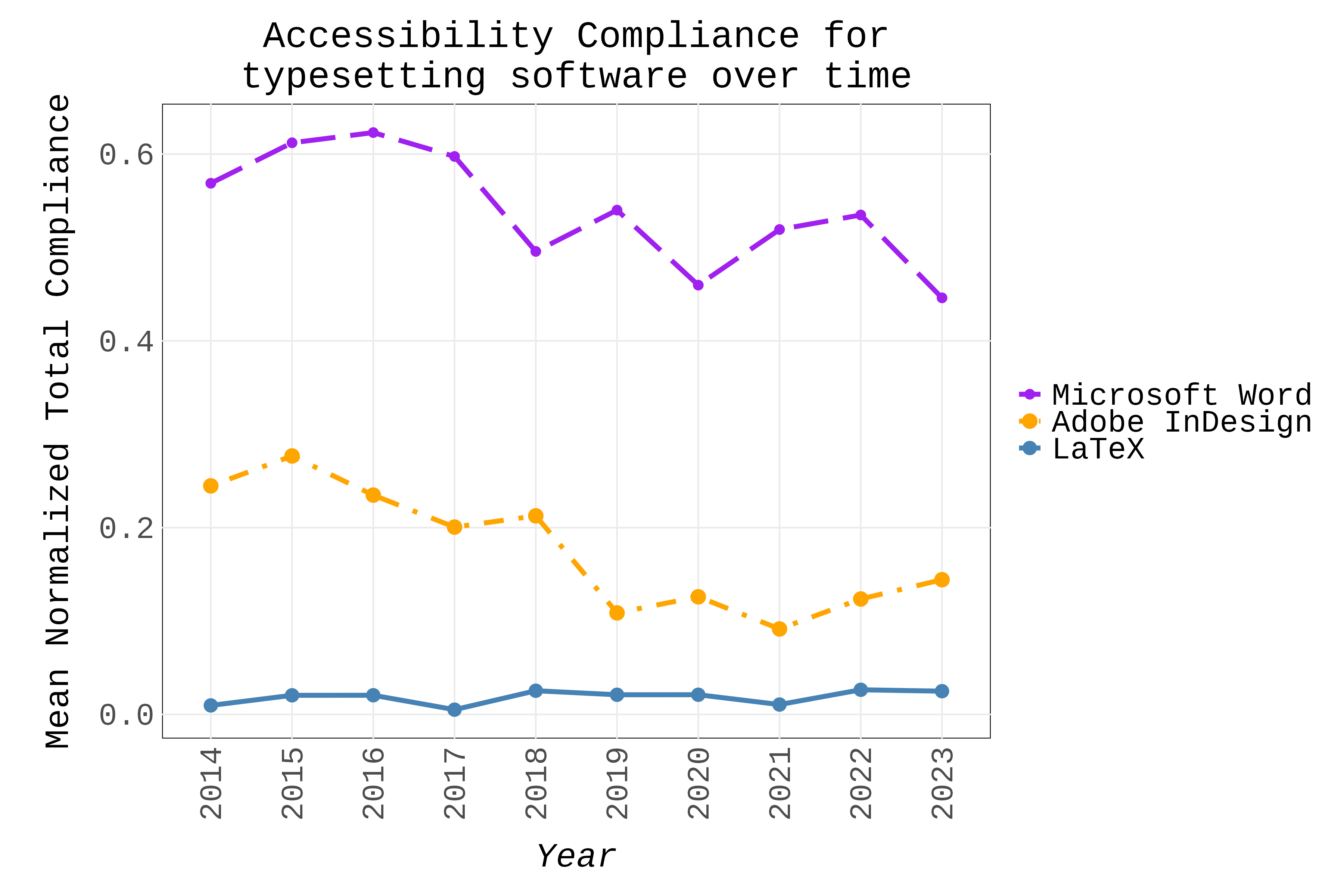}
    \caption{
  Mean normalized accessibility compliance by select PDF creation platforms over time (2014-2023). While accessibility compliance for PDFs created using LaTeX remains relatively stable (and low), there is a significant decline in mean normalized total compliance for PDFs created with Microsoft Word and Adobe InDesign after 2017.} 
  % Following this, their compliance values remain increase slightly.
  
  \label{fig:compliance_software_time}
  \Description{A line plot shows the mean normalized total compliance of scholarly PDFs from 2014 to 2023 for three PDF creation platforms: Adobe InDesign, LaTeX, and Microsoft Word. The x-axis represents the years, and the y-axis represents the proportion of PDFs in our sample. Accessibility compliance for PDFs from LaTeX remains stable but low. In contrast, there is a noticeable decline in compliance for PDFs created with Microsoft Word and Adobe InDesign after 2017.}
\end{figure}

%% file: figtabs/fig_compliance_over_models.tex
\begin{figure}[t!]
  \centering
    \includegraphics[width=0.8\linewidth]{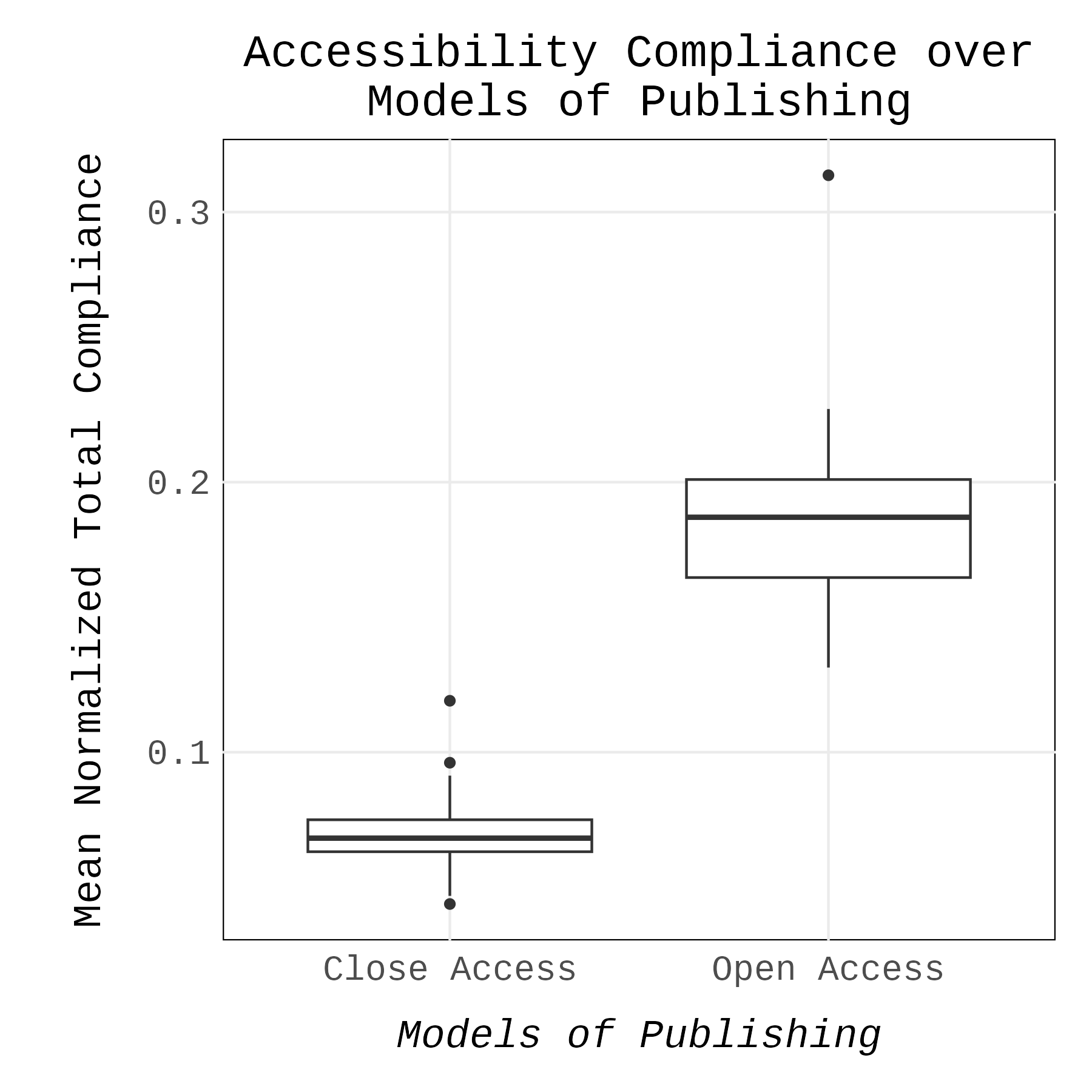}
  \caption{Accessibility compliance for the two models of publishing (open vs. closed). The median of mean normalized total compliance across all fields of study for open-access papers is 0.18 (IQR = 0.04), significantly higher than the 0.07 (IQR = 0.01) observed for closed-access
papers. This demonstrates that open-access papers have
higher accessibility compliance compared to closed access ones. As per the Mann-Whitney U test, there is a significant difference in the distribution of mean normalized total compliance
scores between the two access types (Z = 24.76, p < 0.0001).}
  \label{fig:compliance-over-models}
  \Description{A box plot shows the mean normalized total compliance for open versus closed access publishing models. The median compliance for open-access papers is 0.18 (interquartile range, IQR = 0.04), which is significantly higher than 0.07 (IQR = 0.01) for closed-access papers. This indicates that open-access papers have higher accessibility compliance compared to closed-access papers.}
\end{figure}

%% file: figtabs/tab_checker_comparison.tex
\begin{table*}[t!]
\small
    \centering
    \begin{tabularx}{\linewidth}{L{15mm}L{34mm}L{62mm}L{54mm}}
        \toprule
        \textbf{Criterion} & \textbf{Corresponding Guidelines} & \textbf{Adobe Checker} & \textbf{PAC3} \\
        \midrule
        Default Language & PDF/UA; WCAG 2.1 guideline 3.1.1 (Language of Page) & 
        % Identifies the default language set in a PDF document effectively
        Effective; false positives are infrequent due to integration with Adobe Acrobat, which handles metadata well (0.78\% false positives) & Incorrectly flags documents as missing the language tag, possibly due to less sophisticated recognition of metadata (17.05\% false positives) \\
        \midrule
        Appropriate Nesting & PDF/UA; WCAG 2.1 guideline 1.3.1 (Info and Relationships) & Identifies heading structures accurately, but sometimes fails to detect nested structures in documents with complex layouts (4.8\% false positives) 
        & Tends to over-report compliance, suggesting leniency in recognizing heading structures (75\% false negatives) \\
        \midrule
        Tagged PDF & PDF/UA; WCAG 2.1 does not specifically address PDF tags, but relevant as per guideline 4.1.2 (Name, Role, Value) & Marks documents as compliant, despite having several untagged elements (11.1\% false negatives); overlooks untagged nested elements due to their complex structure, especially tables (1.1\% false negatives) & More rigorous in assessing tagged elements; however, incorrectly interprets whitespace or decorative elements as needing tags (91.8\% false positives)\\
        \midrule
        Table Headers & PDF/UA; WCAG 2.1 guideline 1.3.1 (Info and Relationships) & Misidentifies non-table elements as tables and flags well-structured but complex tables, despite having header tags (34.2\% false positives) & Usually marks tables as compliant and misidentifies styled text as table headers (85.7\% false negatives) \\ 
        \midrule
        Tab Order & PDF/UA does not explicitly address tab order but requires that the document’s navigation be logical and consistent; WCAG 2.1 guideline 2.4.3 (Focus Order) & Mostly accurate and checks tab order to match document structure, but does not assess element-level reading order within pages, which overlooks finer navigation issues & Focuses on keyboard navigation by checking the focus order rather than tab order, which could miss issues related to visual document flow \\
        \midrule
        Alt-Text & PDF/UA; WCAG 2.1 guideline 1.1.1 (Non-text Content) & Checks for the presence of alt-text but does not assess the quality or appropriateness of the descriptions, which could lead to non-descriptive alt-text passing the check. Sometimes, incorrectly flags decorative images that do not require alt-text (10\% false positives) & Similarly checks for alt-text presence without assessing quality; also incorrectly flags decorative images and links that do not require alt-text (10\% false positives)
        \\ 
        % \hline \\ [-2mm]
        % General Comments & Checks for the presence of Alt-text but does not assess the quality or appropriateness of the descriptions, which could lead to non-descriptive alt-text passing the check. Sometimes, incorrectly flags decorative images that do not require alt text (false positives) & Similarly checks for Alt-text presence without assessing the quality
        % \\ [0.6mm]
        % \hline \\ [-2mm]
        % \textbf{\textit{Our analysis}} & \numpdfs & Venues across various fields of study\\
        \bottomrule \\ [-2mm]
    \end{tabularx}
    \caption{Comparative analysis of the Adobe Checker and PAC3, detailing their performance differences when evaluating PDF accessibility. ‘X\% false positives’ refers to the percent of papers reviewed where the tool incorrectly identified a criterion as non-compliant; the denominator is the number of actually compliant documents. Conversely, ‘X\% false negatives’ refers to the percent of papers reviewed where the tool overlooks actual accessibility issues, incorrectly marking the document as compliant; the denominator is the number of non-compliant documents.
    }
    % \Description{
% Prior work, PDFs analyzed, Venues, Year, Accessibility checker 
% Brady et al. [7], 1811, CHI, ASSETS and W4A, 2011--2014, PDFA Inspector 
% Lazar et al. [23], 465 + 32, CHI and ASSETS, 2014--2015, Adobe Acrobat Action Wizard 
% Ribera et al. [40], 59, DSAI, 2016, Adobe PDF Accessibility Checker 2.0 
% Nganji [33], 200, Disability & Society, Journal of Developmental and Physical Disabilities, Journal of Learning Disabilities, and Research in Developmental Disabilities, 2009--2013, Adobe PDF Accessibility Checker 1.3
% Our analysis, 11397, Venues across various fields of study, 2010--2019, Adobe Acrobat Accessibility Plug-in Version 21.001.20145 
    % }
\label{tab:checker_comparison}
\end{table*}

%% file: 4_discussion.tex
\section{Discussion and Recommendations}
\label{sec:discussion}
% Our findings build upon and extend the existing literature on digital accessibility in academic publishing. Previous research has often been limited in scope, focusing on specific fields or publication venues known for their proximity to the accessible computing community. Our study broadens this scope significantly, exploring accessibility trends across a diverse range of disciplines and publication types, over the past decade.
Our analysis reveals a stark landscape of PDF accessibility within academic publishing. We discovered that a large majority of papers in our sample (74.9\%) had zero compliance with accessibility criteria across the spectrum of academic fields, with only a small fraction of papers (3.2\%) fully compliant. Among the assessed criteria, Default Language was the most commonly met, reflecting the minimal effort required for compliance in this area. Conversely, Alt-Text for figures, which necessitates deliberate author input, was the criterion least likely to be met. This underscores a broader issue: the gap between the potential for accessibility and its actual implementation by authors and publishers.

% Significant differences in accessibility compliance were observed across all fields of study, with none of the fields having more than 6.5\% of PDFs satisfying all criteria. However, 
A concerning trend emerged when analyzing the data over time. Before 2019, PDF accessibility was relatively stable or even improving gradually. However, afterwards, we observed a notable and unexpected decline in accessibility compliance, marking a significant shift in the landscape of scholarly PDF accessibility. What went wrong? 
% This decline can be attributed to several factors, some of which we have explored in our analysis. 
Our findings suggest a correlation between open-access papers and higher accessibility compliance, possibly due to the motivation for broader dissemination associated with the open publishing model, encouraging authors to implement accessibility standards more diligently. However, it is among open access papers that we observe the largest drop in accessibility compliance around 2019. We attribute this to several interrelated developments in the publishing ecosystem.

The period after 2019 witnessed a significant increase in the volume of academic publications, especially open access publications. This was driven in part by Plan S,\footnote{Please visit \href{https://www.coalition-s.org/}{https://www.coalition-s.org/} for more detailed information on Plan S} which mandates that scientific publications from publicly funded research be published in compliant Open Access journals or platforms by 2021. 
% As a result, publishers and authors may have expedited the publication process, which could have compromised accessibility standards. Simultaneously, t
The push towards open access may have introduced financial pressures on publishers to reduce costs, which could have resulted in cuts to resources allocated for ensuring accessibility.
Additionally, the shift towards more digital and rapid publishing methods like preprints and online-first publications prioritized speed and broad access to information over compliance with detailed accessibility standards. This trend was further accelerated by the COVID-19 pandemic, which heightened the demand for rapid communication of research findings crucial to public health responses. Indeed, around 16\% of papers in our dataset published after 2019 include COVID-related keywords. However, accessibility changes among COVID-related papers alone do not account for the severity of the drop in accessibility compliance, which implies more systemic and perhaps persistent changes in scholarly publishing.

We also uncovered a strong association between the platform used to create a document and its accessibility compliance. Microsoft Word was found to produce the most accessibility-compliant PDFs, while those produced by XPP, Arbortext APP, and LaTeX had the lowest compliance, underscoring the significant impact of creation platform choice on accessibility. Microsoft has made investments in enhancing the accessibility of their Office 365 Suite,\footnote{\href{https://www.microsoft.com/en-us/accessibility/microsoft-365}{https://www.microsoft.com/en-us/accessibility/microsoft-365}} integrating features like live captions and subtitles in PowerPoint in 2020, which reflects a growing awareness and prioritization of accessibility concerns during document creation. However, the 2019 update of Microsoft Office and subsequent updates introduced AI-driven tools,\footnote{See \href{https://www.microsoft.com/en-us/microsoft-365/blog/2019/06/18/powerpoint-ai-upgrade-designer-major-milestone-1-billion-slides/}{https://www.microsoft.com/en-us/microsoft/blog/2019/powerpoint-ai-upgrade-designer} and \href{https://venturebeat.com/ai/6-ai-features-microsoft-added-to-office-in-2019/}{https://venturebeat.com/ai/features-office-2019/}} aimed at boosting user productivity and enhancing user experience, such as AI in Word offering writing suggestions and PowerPoint suggesting design layouts. While these innovations offered significant benefits, they did not fully integrate necessary accessibility features, which may have contributed to the observed decline in accessibility compliance. For instance, advanced formatting features disrupt the tab order necessary for screen readers. We encourage developers of typesetting and publishing software to prioritize accessibility in their development process to address these issues.

% \lucy{Again, I think the framing for this finding is incorrect. The surprising thing is: before 2019, we thought document accessibility was fairly stable or somewhat improving, and then we saw a massive drop in accessibility around 2018-2019 leading to the current status. We've gone from generally more accessible PDFs to generally less accessible PDFs in science, so what are we doing wrong? Clearly, the policies that we've implemented have not been working. So what should be focus on instead?} Significant differences in accessibility compliance were observed across all fields of study, with none of the fields achieving more than 6.5\% of PDFs satisfying all criteria. Furthermore, accessibility compliance remained relatively stable in recent years, following a notable decline between 2017 and 2020. 

While technological advancements are crucial, we caution that not all aspects of PDF accessibility can be automated through software alone, and their limitations must be recognized. Software may be able to effectively infer reading order and detect structural elements like headings and table headers, which are feasible tasks for visual layout-enhanced language models \citep{Shen2022VILA, Xu2021LayoutLMv2MP} or vision-language models like GPT-4o \citep{GPT4o}. However, certain aspects of PDF accessibility require a deeper level of understanding and contextual decision-making that models alone may struggle to provide. For instance, creating meaningful alt-text for figures demands an understanding of the author's intent and the document context, which is significantly more challenging to automate \citep{Singh_2024}. Authors therefore continue to play an indispensable role in ensuring the accessibility of documents. 
% They are crucial for implementing meaningful alt-Text, ensuring logical reading orders, and correctly using headings, table headers, and structural elements to enhance document navigability and readability. 

% While advancements in software are crucial, we caution that not all aspects of PDF accessibility can be automated through software alone. Authors play an indispensable role, especially in creating meaningful Alt-Text for figures, ensuring logical reading orders, or using headings, table headers, and structural elements correctly to enhance document navigability and readability \lucy{there's a big difference between writing meaningful alt text vs understanding the structure of a document. some of the visual layout-enhanced language models do very well on reading order inference, so there's a hierarchy here that you should discuss. automatically inferring reading order should be technically doable, same with automatically detecting headings, table headers, and so on. automatically writing meaningful alt text is much harder because it needs a model to understand the intention of the authors}. These tasks often require contextual understanding and decision-making that software alone cannot achieve. 

Despite the potential for technology to support accessibility efforts, we have not observed any improvements in accessibility compliance, likely because accessibility concerns are considered marginal, and are outside of the awareness of most publishing authors and researchers. Significant changes in the authorial and publication processes are needed to change this status quo, and to increase the accessibility of scientific papers for BLV readers and screen reader users. Outside of the accessibility and HCI publishing communities, the likelihood of rapid improvement is lower, and automated or crowdsourcing solutions may be needed to improve the accessibility of publications in these other domains. Moreover, the inherent complexity of certain academic materials, such as scientific papers that include complex equations and formats, presents additional challenges that require specialized knowledge and tools to make them accessible. Training and guidelines for authors on creating accessible content, especially in technically dense fields, as well as advocacy for the importance of accessibility, are crucial components to success. 

% Additionally, collaboration between accessibility experts and software developers could lead to more sophisticated tools that better address the specific needs of diverse academic disciplines. 
% Publishers should develop and disseminate comprehensive accessibility guidelines that go beyond minimum legal requirements, and provide training sessions and resources on creating accessible content. Possibly, they should enforce accessible document templates or more compliant typesetting software 
% % as their choice of submission and typesetting 
% in their submissions requirements, and conduct regular accessibility audits using both automated tools and manual evaluations by trained professionals or BLV readers. Furthermore, 
% \lucy{this last paragraph didn't seem like it fit in anywhere, since you start the paragraph by talking about collaboration between accessibility experts and software developers, but then go on to talk about what publishers should do to change policies. commenting out for now...}

PDFs have been repeatedly called out as being inaccessible, not only for screen readers, but broadly for reading, especially on mobile and other devices with small screen sizes \citep{NielsenPDFStillUnfit}. Dissociating publishing from PDFs continues to be a good goal for the future. In recent years, alternative publication formats have risen in popularity, such as eLife's dual publication in PDF and HTML,\footnote{\href{https://reviewer.elifesciences.org/author-guide/post}{https://reviewer.elifesciences.org/author-guide/post}} the interactive HTML papers at distill.pub,\footnote{\href{https://distill.pub/}{https://distill.pub/}} or the ACM Digital Library's very own dual publication (PDF and HTML) process,\footnote{\href{https://www.acm.org/publications/authors/submissions}{https://www.acm.org/publications/authors/submissions}} which is now available for many of the ACM's computing conferences and journals. Additionally, arXiv has introduced HTML formats for nearly all of its newly submitted papers \cite{Frankston2024HTMLPO}. Similarly, projects like Paper to HTML \cite{Paper2HTML} can transform scientific papers into more accessible HTML versions, improving readability on mobile devices and accessibility via screen readers. We have no doubt that more such viable alternatives to PDF will arise, and we encourage the community to explore these options when making publishing decisions.

Hence, the substantial variability in accessibility compliance across different fields of study, modes of publication, and creation software highlights the multifaceted challenges and opportunities in improving PDF accessibility in academic publishing. Our results underscore the need for concerted efforts from authors, publishers, software developers, and academic communities to prioritize and enhance the accessibility of scholarly work.  By adopting a holistic approach involving technology, human effort, and systemic changes in publishing practices, we can transform accessibility from an afterthought to a fundamental component of scholarly output.

% For instance, while a tool can flag missing Alt-Text, it cannot determine whether the Alt-Text adequately describes the figure in a way that is meaningful to those who cannot see it. Similarly, ensuring that the document’s reading order is logical and coherent is a task that can benefit from automated checks but ultimately depends on careful review and manual adjustment by the author. 

% Our assessment of alt-text quality for PDFs that passed the Alt-text criteria reveals that the vast majority of alt-text are not meaningful and may be auto-generated. Rather than addressing accessibility in a meaningful way, the increasing presence of this type of auto-generated alt-text may actually increase the difficulty of measuring and benchmarking accessibility improvements in the future.

\section{Limitations}
\label{sec:limitations}
We acknowledge the limitations of our analysis methods. First, our study employed automated accessibility checkers, enabling us to conduct large-scale analysis of over 20K PDFs, but it inherently limited the depth of our accessibility assessments. The Adobe accessibility checker, used predominantly in our research, provides binary outputs (Passed/Failed) that do not fully capture the nuances of document accessibility, particularly in the evaluation of the quality and meaningfulness of alt-text. Where possible, we have performed smaller scale manual inspection to corroborate our main analysis results. Second, since we did not want to rely on a single accessibility checker, which may have biased our results towards the specific standards and capabilities of the tool, we supplemented our analysis with PAC3 for a secondary assessment on a subset of 152 papers. However, the limited scope of this analysis prevents us from generalizing its findings, underscoring the need for 
% employing multiple automated checkers or developing a 
more sophisticated assessment tools that allow for batch processing, integrate a broader range of accessibility criteria, and offer more granular analysis. Addressing these technological and resource limitations in future work could enable more expansive and robust research, potentially leading to more detailed and actionable insights.

We conducted manual assessments with screen readers (NVDA and VoiceOver) on a stratified sample of 133 PDFs, which revealed discrepancies between automated assessments and the actual user experience. Ideally, these manual assessments should have been conducted with the participation of end-users who rely on these technologies daily. Due to resource and time constraints, we were unable to involve end-users directly in this phase, and instead, conducted these ourselves to supplement our existing large-scale analyses. Future studies could enhance the validity of these findings by incorporating end user testing, which would provide invaluable insights into the real-world accessibility of academic PDFs.

% Additionally, the lack of a programmatic API for our accessibility checkers (Adobe checker and PAC3) necessitated significant manual effort, limiting our ability to process larger numbers of documents and recover from errors efficiently. Addressing these technological and resource limitations in future work could enable more expansive and robust research, potentially leading to more detailed and actionable insights.

Another potential limitation is that there may exist multiple versions of each paper, and one of these versions may be more accessible but not indexed by our data source, OpenAlex. We analyzed the most readily available versions as indexed by OpenAlex, which may not always be the most accessible. This could potentially impact our findings, as we cannot guarantee the capturing of the best possible accessibility score for each paper. Future studies should consider ways to identify and analyze all available versions of a paper to provide a more accurate assessment of its accessibility compliance.
Moreover, our sample primarily contained journal publications, with fewer publication types such as preprints, book chapters, reports, and other educational resources. Expanding the scope of future research to these types of documents could yield a more comprehensive understanding of the current state of accessibility across the entire spectrum of academic publishing. 

The observed low accessibility compliance across different fields can be attributed to various decisions made by authors, publishers, societies, and conference organizers, which are often influenced by each field’s unique practices and standards. As a result, it is challenging to derive field-specific recommendations. Future research should examine field-specific practices and their impact on accessibility to enable the development of more targeted interventions.

% \lucy{maybe also: difficulty generalizing our results to field-specific recommendations, since the reason for low accessibility compliance is the result of many individual decisions by authors, publishers, societies, and conference organizers}

\section{Conclusion}
\label{sec:conclusion}
Our study on scholarly PDF accessibility highlights the significant challenges faced by blind and low-vision readers and screen reader users in accessing academic literature. Our analysis across various disciplines reveals low compliance with accessibility criteria, as well as a concerning drop in compliance led by open access publications in recent years. These disparities are influenced by factors such as PDF creation platforms and publishing models, emphasizing the need to improve both the available tools and awareness and practices within the academic community. We advocate for collaborative efforts among authors, publishers, and software developers to prioritize accessibility in academic publishing. Furthermore, our manual assessments underscore the importance of supplementing automated checker results with comprehensive evaluations to ensure a more accurate measure of accessibility. As the digital landscape evolves, the academic community should consider moving beyond PDFs to more accessible formats to ensure that all scholarly work is truly accessible to everyone, including those with disabilities. This shift not only aligns with ethical standards but also broadens the dissemination and impact of scholarly work.

%% file: appendix.tex
% \newpage

% \section{Additional analyses}
% This section contains additional plots highlighting trends and relationships in scholarly document accessibility.

\section{Correlations between accessibility criteria}
\label{sec:criteria_correlation}
\input{figtabs/fig_correlation_matrix}

In Figure~\ref{fig:criteria_correlation_matrix}, we plot a heatmap that displays the Pearson correlation coefficients for various accessibility criteria used in evaluating PDFs. We observe strong correlations between Alt-Text and Table Headers (0.66), Alt-Text and Appropriate Nesting (0.69), Table Headers and Tagged PDF (0.74), Tagged PDF and Appropriate Nesting (0.86), and Table Headers and Appropriate Nesting (0.89). Additionally, we note moderate correlations between Tagged PDF and Default Language (0.45), as well as Default Language and Tab Order (0.53), indicating these combinations frequently appear together in papers. Other correlations range from lower to moderate, as depicted by varying shades from lighter red to white on the heatmap.

\section{Usage of PDF creation platforms in different publishing models}
\label{sec:typesetting_software_over_time}
\input{figtabs/fig_typesetting_over_time}

To examine the trends in PDF creation platform usage within different publishing models, we plot the usage of the seven major PDF creation platforms for open and closed-access papers over time in Figure~\ref{fig:typesetting_software_over_time}. For open-access papers, Adobe InDesign, LaTeX, and XPP were initially the most utilized platforms. Over the years, we observe a significant increase in the use of Springer and Microsoft Word, while LaTeX's usage has continued to grow steadily. In contrast, XPP's usage has significantly declined, becoming almost negligible for open-access publications. Post-2019, there is a notable increase in the usage of Arbortext APP, corresponding with a decline in the use of Adobe InDesign. For closed-access papers, Arbortext APP was the dominant platform early in the decade but experienced a sharp decline subsequently. The use of other creation platforms has either been stable or has shown a gradual decline, suggesting that these platforms may be becoming obsolete or are less preferred compared to newer alternatives.

\section{Correlations between creation platforms and PDF accessibility}
\label{sec:regression_analysis}
\input{figtabs/fig_prop_all_compliance}
We plot the proportion of usage of seven major PDF creation platforms per field of study and the corresponding mean
normalized Total Compliance rates for those fields, in Figure~\ref{fig:prop_all_compliance}. There is a strong positive correlation between the proportion of PDFs created using Microsoft Word and their accessibility compliance ($r = 0.66$, $p < 0.01$, 95\% CI shown) with higher compliance in fields like Medicine, Environmental Science, Business, Political Science, Sociology, Economics, and Psychology, where its usage exceeds 15\%. Elsevier also shows a strong positive correlation ($r = 0.53$, $p < 0.05$, 95\% CI shown) with higher compliance, especially in Political science and Business. Conversely, LaTeX demonstrates a strong negative correlation ($r = -0.74$, $p < 0.001$, 95\% CI shown) with lower compliance, particularly in fields such as Mathematics, Physics, and Computer Science, where it is used similarly extensively. Correlations for other creation platforms (Adobe InDesign, Springer, and XPP) are non-significant ($r = 0.16$, $0.12$, and $-0.06$ respectively, all non-significant).

%% file: figtabs/fig_correlation_matrix.tex
\begin{figure}[h!]
  \centering
    \includegraphics[width=1.0\linewidth,trim={0mm 2mm 0mm 0mm},clip]{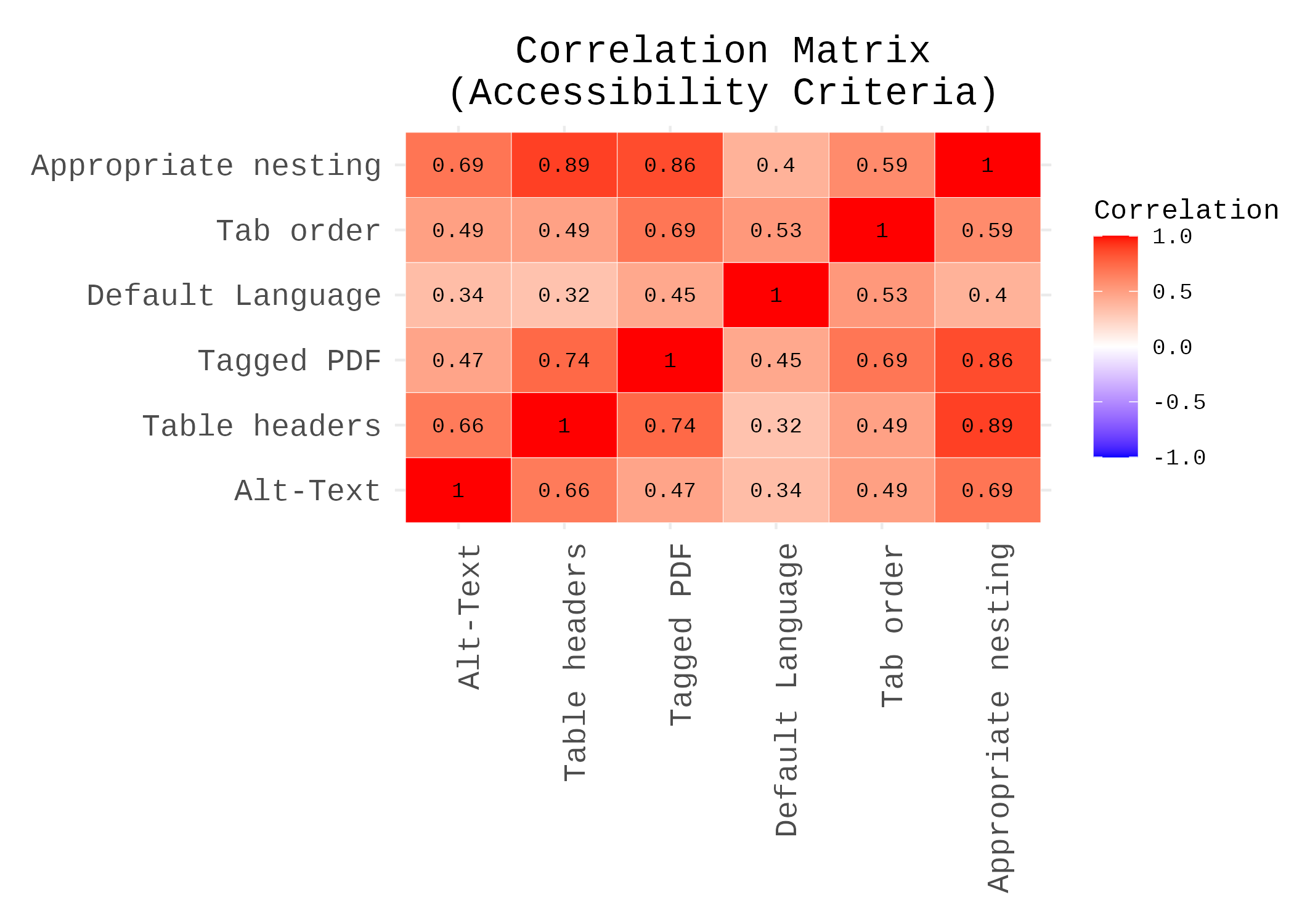}
  \caption{
Pearson correlation coefficients between accessibility criteria. We observe strong positive correlations between several criteria, notably between Appropriate Nesting and Table headers ($r$ = 0.89), Appropriate Nesting and Tagged PDF ($r$ = 0.86), and Tagged PDF and Table Headers ($r$ = 0.74).}
  \label{fig:criteria_correlation_matrix}
  \Description{This heatmap displays the correlation matrix for various accessibility criteria used in evaluating PDFs. The matrix includes six accessibility criteria: Alt-Text, Table headers, Tagged PDF, Default Language, Tab order, and Appropriate nesting. Each cell in the 6x6 grid represents the pearson correlation coefficient between two criteria, with the criteria listed both on the x-axis and y-axis in the same order. Key observations from the figure include the high correlation between Alt-Text and Table headers (r is 0.66), Alt-Text and Appropriate nesting (r is 0.69), Table headers and Tagged PDF (r is 0.74), Tagged PDF and Appropriate nesting (r is 0.86), and Table headers and Appropriate nesting (r is 0.89). Additionally, Tagged PDF and Default Language show a moderate correlation (r is 0.45), as do Default Language and Tab order (r is 0.53). Other correlations vary, with many lower or moderate Pearson correlation coefficient values.}
\end{figure}

%% file: figtabs/fig_typesetting_over_time.tex
\begin{figure}[h!]
  \centering
    \includegraphics[width=0.97\linewidth]{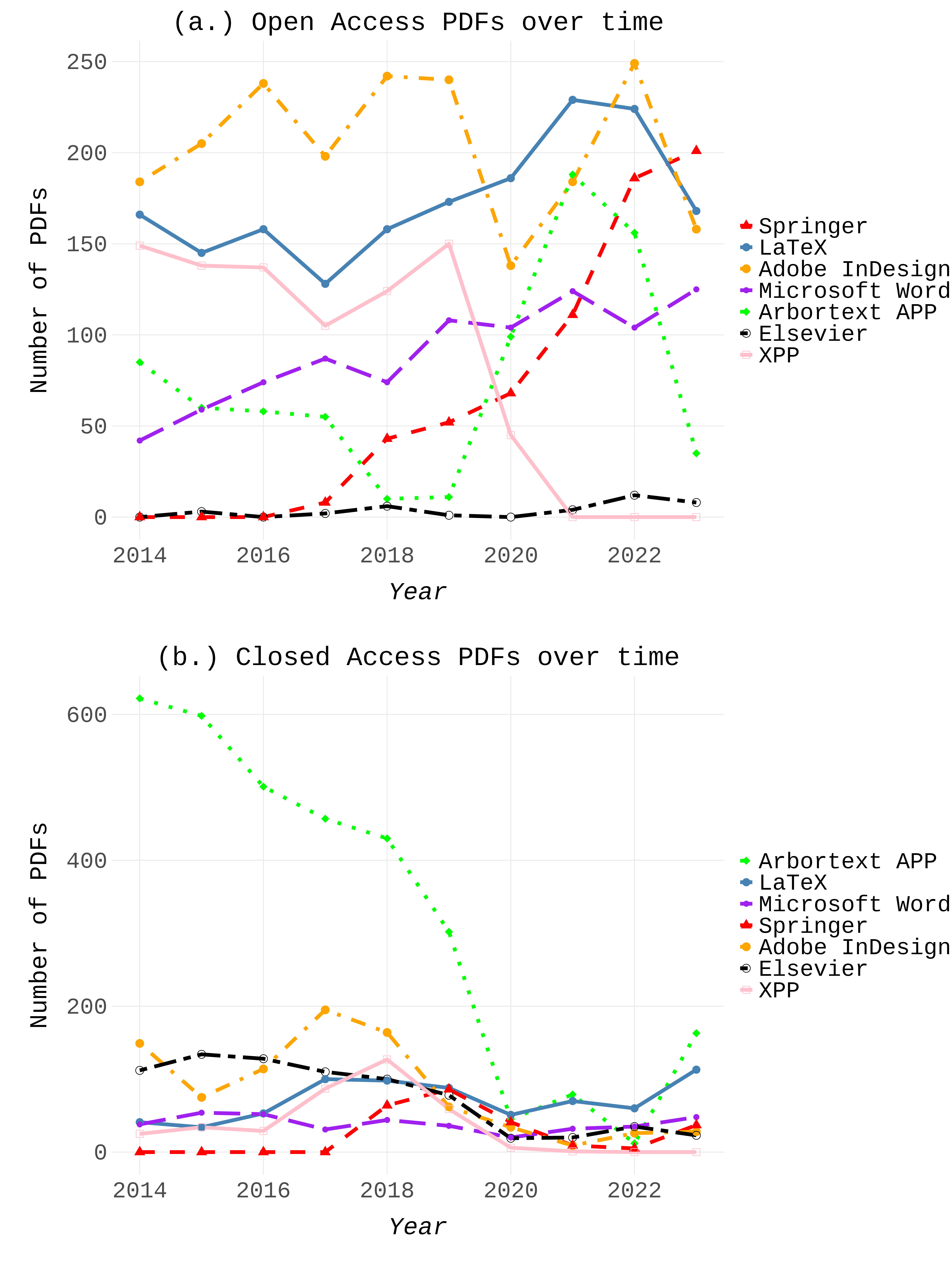}
  \caption{The number of (a) Open Access and (b) Closed Access PDFs created by the seven most common PDF-creation platforms over time.}
  \label{fig:typesetting_software_over_time}
  \Description{The image contains two line plots, each displaying the number of open and closed access PDFs created using different creation platforms over time from 2014 to 2023. In the first plot (for open access PDFs), Adobe InDesign, LaTeX, and XPP were the most utilized platforms early in the decade. Over the years, there has been a significant increase in the use of Springer and Microsoft Word, while LaTeX's usage has continued to grow steadily. In contrast, XPP's usage has significantly declined, becoming almost negligible for open-access publications. Post-2019, there is a notable increase in the usage of Arbortext APP, corresponding with a decline in the use of Adobe InDesign. In the second plot (for closed access PDFs), Arbortext APP was the dominant platform early in the decade but experienced a sharp decline subsequently. However, the use of other creation platforms has either been stable or has shown a gradual decline.}
\end{figure}

%% file: figtabs/fig_prop_all_compliance.tex
\begin{figure*}[ht!]
  \centering
    \includegraphics[width=0.65\linewidth]{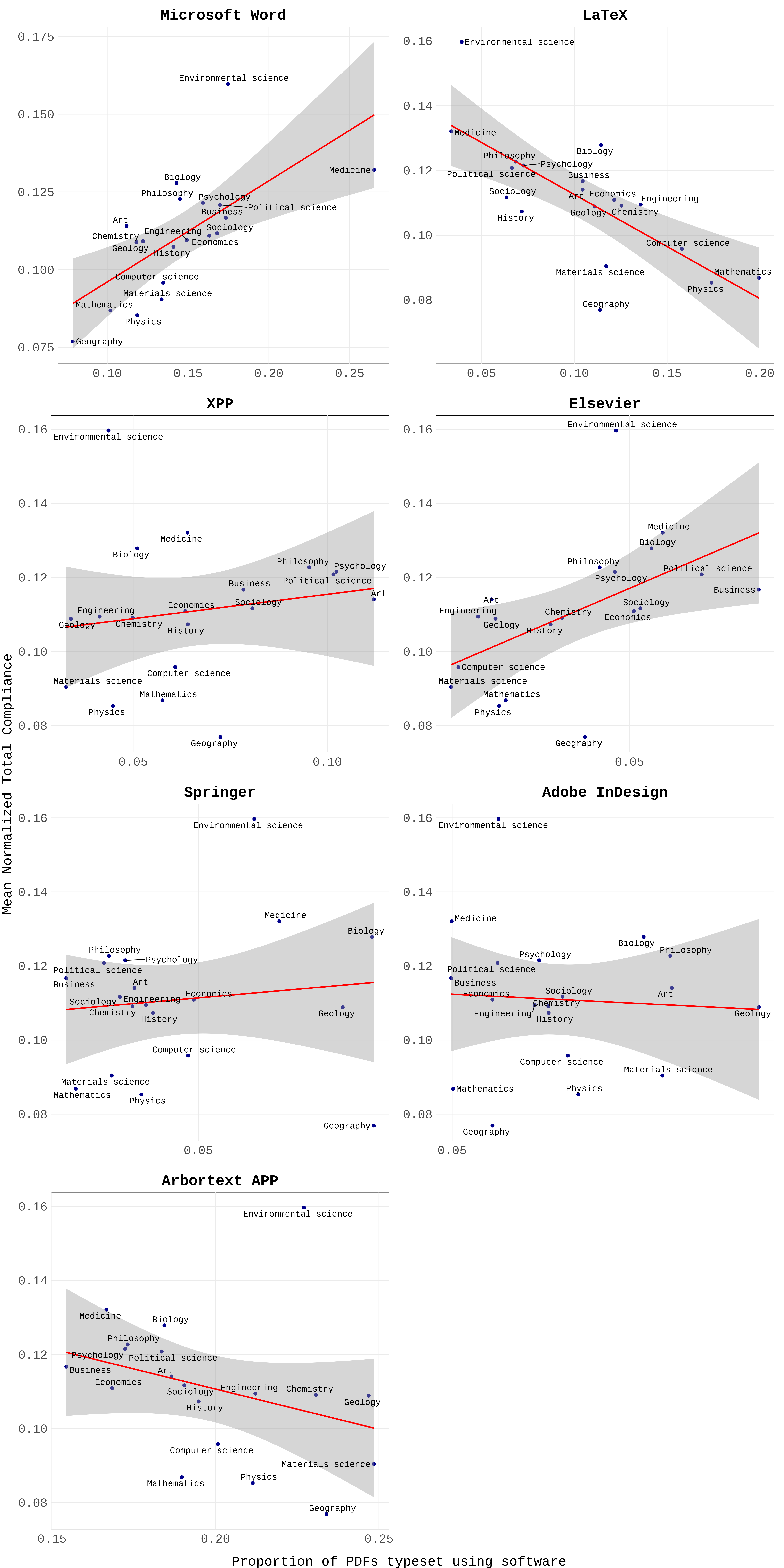}
  \caption{Correlation between PDF Creation Software Usage and Accessibility Compliance by Field of Study. 
  % The figure shows a positive correlation between the proportion of PDFs created using Microsoft Word and their accessibility compliance ($r = 0.66$, $p < 0.01$, 95\% CI shown) with higher compliance in fields like Medicine, Environmental Science, Business, Political Science, Sociology, Economics, and Psychology, where its usage exceeds 15\%. Elsevier also shows a strong positive correlation ($r = 0.53$, $p < 0.05$, 95\% CI shown) with higher compliance, especially in Political science and Business. Conversely, LaTeX demonstrates a strong negative correlation ($r = -0.74$, $p < 0.001$, 95\% CI shown) with lower compliance, particularly in fields such as Mathematics, Physics, and Computer Science, where it is used similarly extensively. Correlations for other creation platforms (Adobe InDesign, Springer, and XPP) are non-significant (r = 0.16, 0.12, and -0.06 respectively, all non-significant)
  }
  \label{fig:prop_all_compliance}
  \Description{Seven scatter plots show the correlation between the proportion of PDFs created by different platforms (Microsoft Word, LaTeX, XPP, Elsevier, Springer, and Adobe InDesign) and their mean normalized total compliance scores across various fields of study. Microsoft Word shows a positive correlation (correlation coefficient r is 0.66 and p less than 0.01) with higher compliance scores in fields like Medicine, Environmental Science, and Psychology. Conversely, LaTeX shows a negative correlation (correlation coefficient r is -0.74 and p less than p < 0.001), with higher usage correlating to lower compliance scores in fields like Mathematics, Physics, and Computer Science. The plot also shows a significant positive correlation for PDFs created using Elsevier (correlation coefficient r is 0.53, p less than 0.05, 95\% Confidence Interval), whereas correlations for PDFs from Adobe InDesign, Springer, and XPP are not significant (correlation coefficients r are 0.16, 0.12, and -0.06, respectively).}
\end{figure*}
% {figures/prop_all_compliance.png}

%% file: main.bbl
%%% -*-BibTeX-*-
%%% Do NOT edit. File created by BibTeX with style
%%% ACM-Reference-Format-Journals [18-Jan-2012].

\begin{thebibliography}{42}

%%% ====================================================================
%%% NOTE TO THE USER: you can override these defaults by providing
%%% customized versions of any of these macros before the \bibliography
%%% command.  Each of them MUST provide its own final punctuation,
%%% except for \shownote{}, \showDOI{}, and \showURL{}.  The latter two
%%% do not use final punctuation, in order to avoid confusing it with
%%% the Web address.
%%%
%%% To suppress output of a particular field, define its macro to expand
%%% to an empty string, or better, \unskip, like this:
%%%
%%% \newcommand{\showDOI}[1]{\unskip}   % LaTeX syntax
%%%
%%% \def \showDOI #1{\unskip}           % plain TeX syntax
%%%
%%% ====================================================================

\ifx \showCODEN    \undefined \def \showCODEN     #1{\unskip}     \fi
\ifx \showDOI      \undefined \def \showDOI       #1{#1}\fi
\ifx \showISBNx    \undefined \def \showISBNx     #1{\unskip}     \fi
\ifx \showISBNxiii \undefined \def \showISBNxiii  #1{\unskip}     \fi
\ifx \showISSN     \undefined \def \showISSN      #1{\unskip}     \fi
\ifx \showLCCN     \undefined \def \showLCCN      #1{\unskip}     \fi
\ifx \shownote     \undefined \def \shownote      #1{#1}          \fi
\ifx \showarticletitle \undefined \def \showarticletitle #1{#1}   \fi
\ifx \showURL      \undefined \def \showURL       {\relax}        \fi
% The following commands are used for tagged output and should be
% invisible to TeX
\providecommand\bibfield[2]{#2}
\providecommand\bibinfo[2]{#2}
\providecommand\natexlab[1]{#1}
\providecommand\showeprint[2][]{arXiv:#2}

\bibitem[Eur({[n.\,d.]})]%
        {European_accessibility_act}
 \bibinfo{year}{[n.\,d.]}\natexlab{}.
\newblock \bibinfo{title}{European accessibility act}.
\newblock
\newblock
\urldef\tempurl%
\url{https://ec.europa.eu/social/main.jsp?catId=1202&langId=en}
\showURL{%
\tempurl}


\bibitem[GPT({[n.\,d.]})]%
        {GPT4o}
 \bibinfo{year}{[n.\,d.]}\natexlab{}.
\newblock \bibinfo{title}{Hello {GPT-4o}}.
\newblock
\newblock
\urldef\tempurl%
\url{https://openai.com/index/hello-gpt-4o/}
\showURL{%
\tempurl}


\bibitem[Angerbauer et~al\mbox{.}(2022)]%
        {Angerbauer2022colorvision}
\bibfield{author}{\bibinfo{person}{Katrin Angerbauer}, \bibinfo{person}{Nils Rodrigues}, \bibinfo{person}{Rene Cutura}, \bibinfo{person}{Seyda \"{O}ney}, \bibinfo{person}{Nelusa Pathmanathan}, \bibinfo{person}{Cristina Morariu}, \bibinfo{person}{Daniel Weiskopf}, {and} \bibinfo{person}{Michael Sedlmair}.} \bibinfo{year}{2022}\natexlab{}.
\newblock \showarticletitle{Accessibility for Color Vision Deficiencies: Challenges and Findings of a Large Scale Study on Paper Figures}.
\newblock \bibinfo{journal}{\emph{Proceedings of the 2022 CHI Conference on Human Factors in Computing Systems}} (\bibinfo{year}{2022}).
\newblock
\showISBNx{9781450391573}
\urldef\tempurl%
\url{https://doi.org/10.1145/3491102.3502133}
\showDOI{\tempurl}


\bibitem[Bigham et~al\mbox{.}(2016)]%
        {Bigham2016AnUT}
\bibfield{author}{\bibinfo{person}{Jeffrey~P. Bigham}, \bibinfo{person}{E. Brady}, \bibinfo{person}{Cole Gleason}, \bibinfo{person}{Anhong Guo}, {and} \bibinfo{person}{D. Shamma}.} \bibinfo{year}{2016}\natexlab{}.
\newblock \showarticletitle{An Uninteresting Tour Through Why Our Research Papers Aren't Accessible}.
\newblock \bibinfo{journal}{\emph{Proceedings of the 2016 CHI Conference Extended Abstracts on Human Factors in Computing Systems}} (\bibinfo{year}{2016}).
\newblock


\bibitem[Brady et~al\mbox{.}(2015)]%
        {Brady2015CreatingAP}
\bibfield{author}{\bibinfo{person}{E. Brady}, \bibinfo{person}{Y. Zhong}, {and} \bibinfo{person}{Jeffrey~P. Bigham}.} \bibinfo{year}{2015}\natexlab{}.
\newblock \showarticletitle{Creating accessible PDFs for conference proceedings}.
\newblock \bibinfo{journal}{\emph{Proceedings of the 12th Web for All Conference}} (\bibinfo{year}{2015}).
\newblock


\bibitem[Brinn et~al\mbox{.}(2022)]%
        {arXiv_2022}
\bibfield{author}{\bibinfo{person}{S. Brinn}, \bibinfo{person}{C. Cameron}, \bibinfo{person}{D. Fielding}, \bibinfo{person}{C. Frankston}, \bibinfo{person}{A. Fromme}, \bibinfo{person}{P. Huang}, \bibinfo{person}{M. Nazzaro}, \bibinfo{person}{S. Orphan}, \bibinfo{person}{S. Sigurdsson}, \bibinfo{person}{R. Tay}, \bibinfo{person}{M. Yang}, {and} \bibinfo{person}{Q. Zhou}.} \bibinfo{year}{2022}\natexlab{}.
\newblock \showarticletitle{A framework for improving the accessibility of research papers on arxiv.org}.
\newblock  (\bibinfo{year}{2022}).
\newblock
\urldef\tempurl%
\url{https://doi.org/10.48550/arxiv.2212.07286}
\showDOI{\tempurl}


\bibitem[Caldwell et~al\mbox{.}(2008)]%
        {Caldwell2008WebCA}
\bibfield{author}{\bibinfo{person}{B. Caldwell}, \bibinfo{person}{M. Cooper}, \bibinfo{person}{Loretta~Guarino Reid}, {and} \bibinfo{person}{G. Vanderheiden}.} \bibinfo{year}{2008}\natexlab{}.
\newblock \showarticletitle{Web Content Accessibility Guidelines (WCAG) 2.0}.
\newblock


\bibitem[Chintalapati et~al\mbox{.}(2022)]%
        {Chintalapati_2022}
\bibfield{author}{\bibinfo{person}{Sanjana~Shivani Chintalapati}, \bibinfo{person}{Jonathan Bragg}, {and} \bibinfo{person}{Lucy~Lu Wang}.} \bibinfo{year}{2022}\natexlab{}.
\newblock \showarticletitle{A Dataset of Alt Texts from HCI Publications: Analyses and Uses Towards Producing More Descriptive Alt Texts of Data Visualizations in Scientific Papers}. In \bibinfo{booktitle}{\emph{Proceedings of the 24th International ACM SIGACCESS Conference on Computers and Accessibility}} \emph{(\bibinfo{series}{ASSETS ’22})}. \bibinfo{publisher}{ACM}.
\newblock
\urldef\tempurl%
\url{https://doi.org/10.1145/3517428.3544796}
\showDOI{\tempurl}


\bibitem[Chisholm et~al\mbox{.}(2001)]%
        {Chisholm2001WebCA}
\bibfield{author}{\bibinfo{person}{W. Chisholm}, \bibinfo{person}{G. Vanderheiden}, {and} \bibinfo{person}{Ian Jacobs}.} \bibinfo{year}{2001}\natexlab{}.
\newblock \showarticletitle{Web content accessibility guidelines 1.0}.
\newblock \bibinfo{journal}{\emph{Interactions}}  \bibinfo{volume}{8} (\bibinfo{year}{2001}), \bibinfo{pages}{35--54}.
\newblock


\bibitem[Darvishy et~al\mbox{.}(2023)]%
        {swiss_2023}
\bibfield{author}{\bibinfo{person}{Alireza Darvishy}, \bibinfo{person}{Rolf Sethe}, \bibinfo{person}{Ines Engler}, \bibinfo{person}{Oriane Pierres}, {and} \bibinfo{person}{Juliet Manning}.} \bibinfo{year}{2023}\natexlab{}.
\newblock \showarticletitle{The state of scientific PDF accessibility in repositories: A survey in Switzerland}.
\newblock \bibinfo{journal}{\emph{Learned Publishing}} \bibinfo{volume}{36}, \bibinfo{number}{4} (\bibinfo{year}{2023}), \bibinfo{pages}{577--584}.
\newblock


\bibitem[Drummer(2012)]%
        {PDF_UA1}
\bibfield{author}{\bibinfo{person}{O. Drummer}.} \bibinfo{year}{2012}\natexlab{}.
\newblock \showarticletitle{{PDF}/{UA} ({ISO} 14289-1) – applying {WCAG} 2.0 principles to the world of {PDF} documents}.
\newblock \bibinfo{journal}{\emph{Proceedings of Computers Helping People with Special Needs}} (\bibinfo{year}{2012}), \bibinfo{pages}{587--594}.
\newblock
\urldef\tempurl%
\url{https://doi.org/10.1007/978-3-642-31522-0_89}
\showDOI{\tempurl}


\bibitem[Drummer and Erle(2012)]%
        {PDF_UA2}
\bibfield{author}{\bibinfo{person}{O. Drummer} {and} \bibinfo{person}{M. Erle}.} \bibinfo{year}{2012}\natexlab{}.
\newblock \showarticletitle{pdf/ua – a new era for document accessibility: understanding, managing and implementing the iso standard pdf/ua (universal accessibility): introduction to the special thematic session}.
\newblock  (\bibinfo{year}{2012}), \bibinfo{pages}{585--586}.
\newblock
\urldef\tempurl%
\url{https://doi.org/10.1007/978-3-642-31522-0_88}
\showDOI{\tempurl}


\bibitem[Frankston et~al\mbox{.}(2024)]%
        {Frankston2024HTMLPO}
\bibfield{author}{\bibinfo{person}{Charles Frankston}, \bibinfo{person}{Jonathan Godfrey}, \bibinfo{person}{Shamsi Brinn}, \bibinfo{person}{Alison Hofer}, {and} \bibinfo{person}{Mark Nazzaro}.} \bibinfo{year}{2024}\natexlab{}.
\newblock \showarticletitle{HTML papers on arXiv - why it is important, and how we made it happen}.
\newblock \bibinfo{journal}{\emph{ArXiv}}  \bibinfo{volume}{abs/2402.08954} (\bibinfo{year}{2024}).
\newblock


\bibitem[i~Graells et~al\mbox{.}(2007)]%
        {Graells2007EstudioDL}
\bibfield{author}{\bibinfo{person}{Miquel~T{\'e}rmens i Graells}, \bibinfo{person}{M.~B. Cerrej{\'o}n}, \bibinfo{person}{M.~D. Boladeras}, \bibinfo{person}{D. Murillo}, \bibinfo{person}{P. Asensio}, {and} \bibinfo{person}{Mireia~Ribera Turr{\'o}}.} \bibinfo{year}{2007}\natexlab{}.
\newblock \showarticletitle{Estudio de la accesibilidad de los documentos cient{\'i}ficos en soporte digital}.
\newblock \bibinfo{journal}{\emph{Revista Espanola De Documentacion Cientifica}}  \bibinfo{volume}{31} (\bibinfo{year}{2007}), \bibinfo{pages}{552--572}.
\newblock


\bibitem[Kelly et~al\mbox{.}(2005)]%
        {Kelly2005}
\bibfield{author}{\bibinfo{person}{B. Kelly}, \bibinfo{person}{D. Sloan}, \bibinfo{person}{L. Phipps}, \bibinfo{person}{H. Petrie}, {and} \bibinfo{person}{F. Hamilton}.} \bibinfo{year}{2005}\natexlab{}.
\newblock \showarticletitle{Forcing standardization or accommodating diversity?: a framework for applying the {WCAG} in the real world}.
\newblock \bibinfo{journal}{\emph{Proceedings of the 2005 International Cross-Disciplinary Workshop on Web Accessibility (W4A)}} (\bibinfo{year}{2005}), \bibinfo{pages}{46--54}.
\newblock
\urldef\tempurl%
\url{https://doi.org/10.1145/1061811.1061820}
\showDOI{\tempurl}


\bibitem[Kruskal and Wallis(1952)]%
        {Kruskal1952UseOR}
\bibfield{author}{\bibinfo{person}{W. Kruskal} {and} \bibinfo{person}{W.~A. Wallis}.} \bibinfo{year}{1952}\natexlab{}.
\newblock \showarticletitle{Use of Ranks in One-Criterion Variance Analysis}.
\newblock \bibinfo{journal}{\emph{J. Amer. Statist. Assoc.}}  \bibinfo{volume}{47} (\bibinfo{year}{1952}), \bibinfo{pages}{583--621}.
\newblock


\bibitem[Lazar et~al\mbox{.}(2017)]%
        {Lazar2017MakingTF}
\bibfield{author}{\bibinfo{person}{J. Lazar}, \bibinfo{person}{E. Churchill}, \bibinfo{person}{T. Grossman}, \bibinfo{person}{G.~V.~D. Veer}, \bibinfo{person}{Philippe~A. Palanque}, \bibinfo{person}{J. Morris}, {and} \bibinfo{person}{Jennifer Mankoff}.} \bibinfo{year}{2017}\natexlab{}.
\newblock \showarticletitle{Making the field of computing more inclusive}.
\newblock \bibinfo{journal}{\emph{Commun. ACM}}  \bibinfo{volume}{60} (\bibinfo{year}{2017}), \bibinfo{pages}{50 -- 59}.
\newblock


\bibitem[Lundgard and Satyanarayan(2022)]%
        {Lundgard2022AccessibleVV}
\bibfield{author}{\bibinfo{person}{Alan Lundgard} {and} \bibinfo{person}{Arvind Satyanarayan}.} \bibinfo{year}{2022}\natexlab{}.
\newblock \showarticletitle{Accessible Visualization via Natural Language Descriptions: A Four-Level Model of Semantic Content}.
\newblock \bibinfo{journal}{\emph{IEEE Transactions on Visualization and Computer Graphics}}  \bibinfo{volume}{28} (\bibinfo{year}{2022}), \bibinfo{pages}{1073--1083}.
\newblock


\bibitem[Mack et~al\mbox{.}(2021a)]%
        {Mack_2021}
\bibfield{author}{\bibinfo{person}{Kelly Mack}, \bibinfo{person}{Edward Cutrell}, \bibinfo{person}{Bongshin Lee}, {and} \bibinfo{person}{Meredith~Ringel Morris}.} \bibinfo{year}{2021}\natexlab{a}.
\newblock \showarticletitle{Designing Tools for High-Quality Alt Text Authoring}. In \bibinfo{booktitle}{\emph{Proceedings of the 23rd International ACM SIGACCESS Conference on Computers and Accessibility}} \emph{(\bibinfo{series}{ASSETS '21})}. Article \bibinfo{articleno}{23}, \bibinfo{numpages}{14}~pages.
\newblock
\urldef\tempurl%
\url{https://doi.org/10.1145/3441852.3471207}
\showDOI{\tempurl}


\bibitem[Mack et~al\mbox{.}(2021b)]%
        {Mack2021DesigningTF}
\bibfield{author}{\bibinfo{person}{Kelly~M. Mack}, \bibinfo{person}{Edward Cutrell}, \bibinfo{person}{Bongshin Lee}, {and} \bibinfo{person}{Meredith~Ringel Morris}.} \bibinfo{year}{2021}\natexlab{b}.
\newblock \showarticletitle{Designing Tools for High-Quality Alt Text Authoring}.
\newblock \bibinfo{journal}{\emph{The 23rd International ACM SIGACCESS Conference on Computers and Accessibility}} (\bibinfo{year}{2021}).
\newblock


\bibitem[Menzies et~al\mbox{.}(2022)]%
        {AuthorReflections2022}
\bibfield{author}{\bibinfo{person}{R. Menzies}, \bibinfo{person}{G. Tigwell}, {and} \bibinfo{person}{M. Crabb}.} \bibinfo{year}{2022}\natexlab{}.
\newblock \showarticletitle{author reflections on creating accessible academic papers}.
\newblock \bibinfo{journal}{\emph{Acm Transactions on accessible Computing}}  \bibinfo{volume}{15} (\bibinfo{year}{2022}), \bibinfo{pages}{1--36}.
\newblock
Issue 4.
\urldef\tempurl%
\url{https://doi.org/10.1145/3546195}
\showDOI{\tempurl}


\bibitem[Nganji(2015)]%
        {Nganji2015ThePD}
\bibfield{author}{\bibinfo{person}{J. Nganji}.} \bibinfo{year}{2015}\natexlab{}.
\newblock \showarticletitle{The Portable Document Format (PDF) accessibility practice of four journal publishers}.
\newblock \bibinfo{journal}{\emph{Library \& Information Science Research}}  \bibinfo{volume}{37} (\bibinfo{year}{2015}), \bibinfo{pages}{254--262}.
\newblock


\bibitem[Nganji(2018)]%
        {Nganji2018ThePD}
\bibfield{author}{\bibinfo{person}{J. Nganji}.} \bibinfo{year}{2018}\natexlab{}.
\newblock \showarticletitle{An assessment of the accessibility of PDF versions of selected journal articles published in a {WCAG} 2.0 era (2014–2018)}.
\newblock \bibinfo{journal}{\emph{Learned Publishing}}  \bibinfo{volume}{31} (\bibinfo{year}{2018}), \bibinfo{pages}{391--401}.
\newblock
Issue 4.
\urldef\tempurl%
\url{https://doi.org/10.1002/leap.1197}
\showDOI{\tempurl}


\bibitem[Nielsen and Kaley(2020)]%
        {NielsenPDFStillUnfit}
\bibfield{author}{\bibinfo{person}{Jakob Nielsen} {and} \bibinfo{person}{Anna Kaley}.} \bibinfo{year}{2020}\natexlab{}.
\newblock \bibinfo{title}{{PDF}: Still Unfit for Human Consumption, 20 Years Later}.
\newblock \bibinfo{howpublished}{\url{https://www.nngroup.com/articles/pdf-unfit-for-human-consumption/}}.
\newblock
\newblock
\shownote{Accessed: 2021-01-31}.


\bibitem[Pradhan et~al\mbox{.}(2022)]%
        {Pradhan_2022}
\bibfield{author}{\bibinfo{person}{Debashish Pradhan}, \bibinfo{person}{Tripti Rajput}, \bibinfo{person}{Aravind~Jembu Rajkumar}, \bibinfo{person}{Jonathan Lazar}, \bibinfo{person}{Rajiv Jain}, \bibinfo{person}{Vlad~I. Morariu}, {and} \bibinfo{person}{Varun Manjunatha}.} \bibinfo{year}{2022}\natexlab{}.
\newblock \showarticletitle{Development and Evaluation of a Tool for Assisting Content Creators in Making PDF Files More Accessible}.
\newblock \bibinfo{journal}{\emph{ACM Trans. Access. Comput.}} \bibinfo{volume}{15}, \bibinfo{number}{1}, Article \bibinfo{articleno}{3} (\bibinfo{date}{mar} \bibinfo{year}{2022}), \bibinfo{numpages}{52}~pages.
\newblock
\showISSN{1936-7228}
\urldef\tempurl%
\url{https://doi.org/10.1145/3507661}
\showDOI{\tempurl}


\bibitem[Priem et~al\mbox{.}(2022)]%
        {priem2022openalexfullyopenindexscholarly}
\bibfield{author}{\bibinfo{person}{Jason Priem}, \bibinfo{person}{Heather Piwowar}, {and} \bibinfo{person}{Richard Orr}.} \bibinfo{year}{2022}\natexlab{}.
\newblock \bibinfo{title}{OpenAlex: A fully-open index of scholarly works, authors, venues, institutions, and concepts}.
\newblock
\newblock
\showeprint[arxiv]{2205.01833}~[cs.DL]
\urldef\tempurl%
\url{https://arxiv.org/abs/2205.01833}
\showURL{%
\tempurl}


\bibitem[Rajkumar et~al\mbox{.}(2020)]%
        {Rajkumar2020PDFAO}
\bibfield{author}{\bibinfo{person}{A.~J. Rajkumar}, \bibinfo{person}{J. Lazar}, \bibinfo{person}{J.~B. Jordan}, \bibinfo{person}{Alireza Darvishy}, {and} \bibinfo{person}{H. Hutter}.} \bibinfo{year}{2020}\natexlab{}.
\newblock \showarticletitle{PDF Accessibility of Research Papers: What Tools are Needed for Assessment and Remediation?}. In \bibinfo{booktitle}{\emph{HICSS}}.
\newblock


\bibitem[Ribera(2008)]%
        {Ribera2008AreAccessible}
\bibfield{author}{\bibinfo{person}{M. Ribera}.} \bibinfo{year}{2008}\natexlab{}.
\newblock \showarticletitle{Are {PDF} Documents Accessible?}
\newblock \bibinfo{journal}{\emph{Information Technology and Libraries}}  \bibinfo{volume}{27} (\bibinfo{year}{2008}), \bibinfo{pages}{25--43}.
\newblock
Issue 3.
\urldef\tempurl%
\url{https://doi.org/10.6017/ital.v27i3.3246}
\showDOI{\tempurl}


\bibitem[Ribera et~al\mbox{.}(2019)]%
        {Ribera2019PublishingAP}
\bibfield{author}{\bibinfo{person}{M. Ribera}, \bibinfo{person}{R. Pozzobon}, {and} \bibinfo{person}{S. Sayago}.} \bibinfo{year}{2019}\natexlab{}.
\newblock \showarticletitle{Publishing accessible proceedings: the DSAI 2016 case study}.
\newblock \bibinfo{journal}{\emph{Universal Access in the Information Society}} (\bibinfo{year}{2019}), \bibinfo{pages}{1--13}.
\newblock


\bibitem[Schmitt-Koopmann et~al\mbox{.}(2022)]%
        {Schmitt-Koopmann_2022}
\bibfield{author}{\bibinfo{person}{F. Schmitt-Koopmann}, \bibinfo{person}{E. Huang}, {and} \bibinfo{person}{A. Darvishy}.} \bibinfo{year}{2022}\natexlab{}.
\newblock \showarticletitle{Accessible {PDFs}: Applying Artificial Intelligence for Automated Remediation of {STEM} {PDFs}}.
\newblock \bibinfo{journal}{\emph{Proceedings of the 24th International ACM SIGACCESS Conference on Computers and Accessibility}} (\bibinfo{year}{2022}), \bibinfo{pages}{1--6}.
\newblock
\urldef\tempurl%
\url{https://doi.org/10.1145/3517428.3550407}
\showDOI{\tempurl}


\bibitem[Shen et~al\mbox{.}(2022)]%
        {Shen2022VILA}
\bibfield{author}{\bibinfo{person}{Zejiang Shen}, \bibinfo{person}{Kyle Lo}, \bibinfo{person}{Lucy~Lu Wang}, \bibinfo{person}{Bailey Kuehl}, \bibinfo{person}{Daniel~S. Weld}, {and} \bibinfo{person}{Doug Downey}.} \bibinfo{year}{2022}\natexlab{}.
\newblock \showarticletitle{VILA: Improving Structured Content Extraction from Scientific PDFs Using Visual Layout Groups}.
\newblock \bibinfo{journal}{\emph{Transactions of the Association for Computational Linguistics (TACL)}} (\bibinfo{year}{2022}).
\newblock


\bibitem[Shen et~al\mbox{.}(2018)]%
        {Shen2018AWS}
\bibfield{author}{\bibinfo{person}{Zhihong Shen}, \bibinfo{person}{Hao Ma}, {and} \bibinfo{person}{Kuansan Wang}.} \bibinfo{year}{2018}\natexlab{}.
\newblock \showarticletitle{A Web-scale system for scientific knowledge exploration}. In \bibinfo{booktitle}{\emph{Proceedings of {ACL} 2018, System Demonstrations}}. \bibinfo{publisher}{Association for Computational Linguistics}, \bibinfo{address}{Melbourne, Australia}, \bibinfo{pages}{87--92}.
\newblock
\urldef\tempurl%
\url{https://doi.org/10.18653/v1/P18-4015}
\showDOI{\tempurl}


\bibitem[Singh et~al\mbox{.}(2024)]%
        {Singh_2024}
\bibfield{author}{\bibinfo{person}{Nikhil Singh}, \bibinfo{person}{Lucy~Lu Wang}, {and} \bibinfo{person}{Jonathan Bragg}.} \bibinfo{year}{2024}\natexlab{}.
\newblock \showarticletitle{FigurA11y: AI Assistance for Writing Scientific Alt Text}. In \bibinfo{booktitle}{\emph{Proceedings of the 29th International Conference on Intelligent User Interfaces}} \emph{(\bibinfo{series}{IUI '24})}. \bibinfo{publisher}{Association for Computing Machinery}, \bibinfo{address}{New York, NY, USA}, \bibinfo{pages}{886–906}.
\newblock
\showISBNx{9798400705083}
\urldef\tempurl%
\url{https://doi.org/10.1145/3640543.3645212}
\showDOI{\tempurl}


\bibitem[Trewin(2014)]%
        {Trewin_2014}
\bibfield{author}{\bibinfo{person}{Shari Trewin}.} \bibinfo{year}{2014}\natexlab{}.
\newblock \bibinfo{title}{Accessible PDF Author Guide}.
\newblock
\newblock
\urldef\tempurl%
\url{https://www.sigaccess.org/welcome-to-sigaccess/resources/accessible-pdf-author-guide/}
\showURL{%
\tempurl}


\bibitem[{Wang} et~al\mbox{.}(2019)]%
        {msr:mag1}
\bibfield{author}{\bibinfo{person}{Kuansan {Wang}}, \bibinfo{person}{Zhihong {Shen}}, \bibinfo{person}{Chiyuan {Huang}}, \bibinfo{person}{Chieh-Han {Wu}}, \bibinfo{person}{Darrin {Eide}}, \bibinfo{person}{Yuxiao {Dong}}, \bibinfo{person}{Junjie {Qian}}, \bibinfo{person}{Anshul {Kanakia}}, \bibinfo{person}{Alvin {Chen}}, {and} \bibinfo{person}{Richard {Rogahn}}.} \bibinfo{year}{2019}\natexlab{}.
\newblock \showarticletitle{A Review of Microsoft Academic Services for Science of Science Studies}.
\newblock \bibinfo{journal}{\emph{Frontiers in Big Data}}  \bibinfo{volume}{2} (\bibinfo{year}{2019}).
\newblock


\bibitem[Wang et~al\mbox{.}(2023)]%
        {Paper2HTML}
\bibfield{author}{\bibinfo{person}{Lucy~Lu Wang}, \bibinfo{person}{Jonathan Bragg}, {and} \bibinfo{person}{Daniel~S. Weld}.} \bibinfo{year}{2023}\natexlab{}.
\newblock \showarticletitle{Paper to HTML: A Publicly Available Web Tool for Converting Scientific Pdfs into Accessible HTML}.
\newblock \bibinfo{journal}{\emph{SIGACCESS Access. Comput.}} \bibinfo{number}{134}, Article \bibinfo{articleno}{1} (\bibinfo{date}{jan} \bibinfo{year}{2023}), \bibinfo{numpages}{1}~pages.
\newblock
\showISSN{1558-2337}
\urldef\tempurl%
\url{https://doi.org/10.1145/3582298.3582299}
\showDOI{\tempurl}


\bibitem[Wang et~al\mbox{.}(2021)]%
        {Wang2021ImprovingTA}
\bibfield{author}{\bibinfo{person}{Lucy~Lu Wang}, \bibinfo{person}{Isabel Cachola}, \bibinfo{person}{Jonathan Bragg}, \bibinfo{person}{Evie (Yu-Yen) Cheng}, \bibinfo{person}{Chelsea~Hess Haupt}, \bibinfo{person}{Matt Latzke}, \bibinfo{person}{Bailey Kuehl}, \bibinfo{person}{Madeleine van Zuylen}, \bibinfo{person}{Linda~M. Wagner}, {and} \bibinfo{person}{Daniel~S. Weld}.} \bibinfo{year}{2021}\natexlab{}.
\newblock \showarticletitle{Improving the Accessibility of Scientific Documents: Current State, User Needs, and a System Solution to Enhance Scientific PDF Accessibility for Blind and Low Vision Users}.
\newblock \bibinfo{journal}{\emph{ArXiv}}  \bibinfo{volume}{abs/2105.00076} (\bibinfo{year}{2021}).
\newblock
\urldef\tempurl%
\url{https://api.semanticscholar.org/CorpusID:233481964}
\showURL{%
\tempurl}


\bibitem[Watson(1990)]%
        {ADA1990}
\bibfield{author}{\bibinfo{person}{P.~G. Watson}.} \bibinfo{year}{1990}\natexlab{}.
\newblock \showarticletitle{The americans with disabilities act: more rights for people with disabilities}.
\newblock \bibinfo{journal}{\emph{Rehabilitation Nursing}}  \bibinfo{volume}{15} (\bibinfo{year}{1990}), \bibinfo{pages}{325--328}.
\newblock
Issue 6.
\urldef\tempurl%
\url{https://doi.org/10.1002/j.2048-7940.1990.tb01505.x}
\showDOI{\tempurl}


\bibitem[Williams et~al\mbox{.}(2022)]%
        {Williams}
\bibfield{author}{\bibinfo{person}{Candace Williams}, \bibinfo{person}{Lilian de Greef}, \bibinfo{person}{Ed Harris}, \bibinfo{person}{Leah Findlater}, \bibinfo{person}{Amy Pavel}, {and} \bibinfo{person}{Cynthia Bennett}.} \bibinfo{year}{2022}\natexlab{}.
\newblock \showarticletitle{Toward supporting quality alt text in computing publications}. In \bibinfo{booktitle}{\emph{Proceedings of the 19th International Web for All Conference}} (Lyon, France) \emph{(\bibinfo{series}{W4A '22})}. \bibinfo{publisher}{Association for Computing Machinery}, \bibinfo{address}{New York, NY, USA}, Article \bibinfo{articleno}{20}, \bibinfo{numpages}{12}~pages.
\newblock
\showISBNx{9781450391702}
\urldef\tempurl%
\url{https://doi.org/10.1145/3493612.3520449}
\showDOI{\tempurl}


\bibitem[Xu et~al\mbox{.}(2021)]%
        {Xu2021LayoutLMv2MP}
\bibfield{author}{\bibinfo{person}{Yang Xu}, \bibinfo{person}{Yiheng Xu}, \bibinfo{person}{Tengchao Lv}, \bibinfo{person}{Lei Cui}, \bibinfo{person}{Furu Wei}, \bibinfo{person}{Guoxin Wang}, \bibinfo{person}{Yijuan Lu}, \bibinfo{person}{Dinei A.~F. Flor{\^e}ncio}, \bibinfo{person}{Cha Zhang}, \bibinfo{person}{Wanxiang Che}, \bibinfo{person}{Min Zhang}, {and} \bibinfo{person}{Lidong Zhou}.} \bibinfo{year}{2021}\natexlab{}.
\newblock \showarticletitle{LayoutLMv2: Multi-modal Pre-training for Visually-Rich Document Understanding}.
\newblock \bibinfo{journal}{\emph{Proceedings of the 59th Annual Meeting of the Association for Computational Linguistics and the 11th International Joint Conference on Natural Language Processing}}  \bibinfo{volume}{abs/2012.14740} (\bibinfo{year}{2021}).
\newblock


\bibitem[Yamada et~al\mbox{.}(2023)]%
        {open_access}
\bibfield{author}{\bibinfo{person}{Y. Yamada}, \bibinfo{person}{A. Nishikawa-Pacher}, {and} \bibinfo{person}{J.~A. T.~d. Silva}.} \bibinfo{year}{2023}\natexlab{}.
\newblock \showarticletitle{Is it open access if registration is required to obtain scientific content?}
\newblock \bibinfo{journal}{\emph{European Science Editing}}  \bibinfo{volume}{49} (\bibinfo{year}{2023}).
\newblock
\urldef\tempurl%
\url{https://doi.org/10.3897/ese.2023.e98101}
\showDOI{\tempurl}


\bibitem[Zulfiqar et~al\mbox{.}(2020)]%
        {Zulfiqar_2020}
\bibfield{author}{\bibinfo{person}{Shaban Zulfiqar}, \bibinfo{person}{Safa Arooj}, \bibinfo{person}{Umar Hayat}, \bibinfo{person}{Suleman Shahid}, {and} \bibinfo{person}{Asim Karim}.} \bibinfo{year}{2020}\natexlab{}.
\newblock \showarticletitle{Automated Generation of Accessible PDF}. In \bibinfo{booktitle}{\emph{Proceedings of the 22nd International ACM SIGACCESS Conference on Computers and Accessibility}} \emph{(\bibinfo{series}{ASSETS '20})}. \bibinfo{publisher}{Association for Computing Machinery}, \bibinfo{address}{New York, NY, USA}, Article \bibinfo{articleno}{91}, \bibinfo{numpages}{3}~pages.
\newblock
\showISBNx{9781450371032}
\urldef\tempurl%
\url{https://doi.org/10.1145/3373625.3418045}
\showDOI{\tempurl}


\end{thebibliography}
